
\input phyzzx.tex
\input tables.tex
\def\msusycol{{M_{SUSY}^{QCD}}}
\def\etal{{\it et al.}}

\def\mtil{\wtil m}
\def\rl{r_{\slepl}}
\def\rr{r_{\slepr}}
\def\lam{\lambda}
\def\gam{\gamma}
\def\mw{m_W}
\def\xw{x_W}
\def\tgut{t_U}
\def\tz{t_Z}
\def\ng{N_g}
\def\msusy{M_{SUSY}}
\def\alphazo{\alpha^0_3(\mz)}
\def\msbar{\overline{MS}}
\def\mplanck{M_{Pl}}
\def\alphaz{\alpha_3(\mz)}
\def\sinsq{\sin^2{\theta_W}}
\def\ibid{{\it ibid.}}

\def\lamt{\lambda_t}
\def\lamb{\lambda_b}

\def\lamtp{\lambda_{\tpr }}
\def\lambp{\lambda_{\bpr}}
\def\lamtaup{\lambda_{\taupr }}

\def\mtp{m_{\tpr }}
\def\mbp{m_{\bpr}}
\def\mnup{m_{\nupr }}
\def\mtaup{m_{\taupr }}

\def\prdj#1{{\it Phys. Rev.} {\bf D{#1}}}
\def\npbj#1{{\it Nucl. Phys.} {\bf B{#1}}}
\def\prlj#1{{\it Phys. Rev. Lett.} {\bf {#1}}}
\def\plbj#1{{\it Phys. Lett.} {\bf B{#1}}}

\def\anti{\overline}
\def\pbi{~{\rm pb}^{-1}}

\def\pb{~{\rm pb}}

\catcode`\@=11 

\def\t1{{\tilde 1}}

\def\gev{\,{\rm GeV}}
\def\tev{\,{\rm TeV}}

\def\wt{\widetilde}
\def\wtil{\widetilde}

\def\rta{\rightarrow}
\def\mhalf{m_{1/2}}
\def\gl{\wt g}
\def\mgl{m_{\gl}}
\def\stop{\wt t}
\def\mstop{m_{\stop}}
\def\sq{\wt q}

\def\msq{m_{\sq}}
\def\sup{\wt u}
\def\sdn{\wt d}
\def\msup{m_{\sup}}
\def\msdn{m_{\sdn}}

\def\supl{\wt u_L}
\def\sdnl{\wt d_L}
\def\msupl{m_{\supl}}
\def\msdnl{m_{\sdnl}}
\def\supr{\wt u_R}
\def\sdnr{\wt d_R}
\def\msupr{m_{\supr}}
\def\msdnr{m_{\sdnr}}

\def\slepl{\wt \ell_L}

\def\mslepl{m_{\slepl}}
\def\slepr{\wt \ell_R}

\def\mslepr{m_{\slepr}}

\def\hl{h^0}
\def\hh{H^0}
\def\ha{A^0}

\def\mhl{m_{\hl}}
\def\mhlmax{\mhl^{\rm max}}
\def\mhh{m_{\hh}}
\def\mha{m_{\ha}}

\def\tanb{\tan\beta}
\def\mt{m_t}
\def\mb{m_b}
\def\mz{m_Z}
\def\mgut{M_U}
\def\taup{\taupr}
\def\tp{\tpr}
\def\bp{\bpr}
\def\bpr{b^\prime}
\def\tpr{t^\prime}
\def\nupr{\nu^\prime}
\def\sbp{\wtil b^\prime}
\def\stp{\wtil t^\prime}
\def\sbpone{\wtil b^\prime_1}

\def\nupr{\nu^\prime}
\def\taupr{\tau^\prime}
\def\mbpr{m_{\bpr}}
\def\mtpr{m_{\tpr}}
\def\mnupr{m_{\nupr}}
\def\mtaupr{m_{\taupr}}

\def\cnone{\wt\chi^0_1}
\def\cntwo{\wt\chi^0_2}
\def\snu{\wt\nu}

\def\msnu{m_{\snu}}

\def\mcnone{m_{\cnone}}
\def\mcntwo{m_{\cntwo}}
\def\h{h}
\def\mh{m_{\h}}
\def\cpone{\wt \chi^+_1}
\def\cmone{\wt \chi^-_1}
\def\cpmone{\wt \chi^{\pm}_1}
\def\mcpmone{m_{\cpmone}}
\def\mcpone{m_{\cpone}}
\def\cptwo{\wt \chi^+_2}

\def\mcptwo{m_{\cptwo}}
\def\stau{\wt \tau}

\def\stauone{\wt \tau_1}

\def\staupone{\wt \taup_1}
\def\mstaupone{m_{\staupone}}
\def\staup{\wt \taup}

\def\nup{\nu^\prime}
\def\snup{\wt \nup}

\def\snupone{\wt \nup_1}
\def\msnupone{m_{\snupone}}
\date{May, 1995}
\Pubnum{$\caps UCD-95-18$\cr}
\titlepage
\baselineskip 0pt
\bigskip
\centerline{\bf A MINIMAL FOUR-FAMILY SUPERGRAVITY MODEL}
\vskip .05in
\centerline{J.~F.~GUNION$^{(a)}$, Douglas W. McKAY$^{(b)}$ and H. POIS$^{(a)}$}
\smallskip
\centerline{(a) \it Davis Institute for High Energy Physics,}
\centerline{\it Department of Physics, U. C. Davis, Davis, CA 95616}
\centerline{(b) \it University of Kansas, Department of Physics and
Astronomy,}
\centerline{ \it Lawrence, Kansas, 66045}
\vskip .1in
\centerline{\bf Abstract}
In this work, we investigate the phenomenology of minimal four-family
MSSM supergravity theories containing an additional
generation ($\tpr,\bpr,\taupr,\nupr$) of heavy fermions along with their
superpartners. We constrain the models by demanding:
gauge coupling constant unification at high energy scales; perturbative values
for all Yukawa couplings for energy scales up to the grand-unification scale;
radiative electroweak (EW) symmetry breaking via
renormalization group evolution down from the grand-unification scale;
a neutral LSP; and consistency with constraints from direct searches for new
particles and precision electroweak data. The perturbative
constraints imply a rather light fourth-family  quark and lepton
spectrum, and $\tanb\lsim 3$.  The lightest CP-even Higgs mass receives
fourth-family loop corrections that can result in as much as a $30\%$
increase over the corresponding three-family mass value.
Significant fourth-family Yukawa coupling contributions to
the evolution of scalar masses lead to unexpected
mass hierarchies among the sparticles. For example, the $\staupone$
is generally the lightest slepton and the lightest squark is the $\wt\bpr_1$.
A significant lower bound is placed on the gluino mass by the
simple requirement that the $\staupone$ not be the LSP.
Sleptons of the first two families are much more massive compared
to the LSP and other neutralinos and charginos than in the three-family
models; in particular, all sleptons belonging to the first three families
could easily lie beyond the reach of a $\sqrt s=500\gev$ $\epem$ collider.
Consistency tests of the RGE equations via mass sum rules and relations
are explored. Relations between slepton masses and gaugino masses
are shown to be very sensitive to the presence of a fourth generation.
The most important near-future experimental probes of the four-family
models are reviewed.
A scenario with $\mt\sim\mw$ and $t\rta {\wt t_1}\cnone$ is shown
to be inconsistent with universal soft-SUSY-breaking boundary conditions.
Full four-family evolution of $\alpha_s$ is shown to lead to a significant
enhancement in inclusive jet and di-jet spectra at Tevatron
energies when all sparticle masses are near their lower bounds.

\baselineskip 0pt

\bigskip
\leftline{\bf 1. Introduction}

Despite the success of the Standard Model (SM), it is almost certainly
incomplete. In particular, the full theory should include a quantum
theory of gravity. However, the huge discrepancy between the
characteristic Planck scale ($\mplanck\simeq 10^{18}\gev$) and
the weak scale, ${\cal O}(\mz)$, is not easily bridged without
encountering problems related to fine-tuning and gauge-hierarchy.
Supersymmetry is currently the only fully quantitative and consistent
resolution to these problems. In particular,
the Minimal Supersymmetric extension of the Standard Model (MSSM)
is an extremely attractive contender for physics
beyond the SM due to its natural prediction that the gauge couplings
all unify at a common grand-unification (GUT) scale, $\mgut$.
The precise particle content of the MSSM (comprising sparticle partners
for all SM particles and exactly two Higgs doublet fields $H_1,H_2$
and their higgsino partners) is crucial to this success.
In addition, the MSSM
provides an attractive explanation for the long proton decay lifetime and
an attractive candidate (the lightest supersymmetric particle, or LSP)
for cold dark matter, easily accommodates the absence of significant
flavor-changing neutral currents (FCNC's), and is completely consistent
with current precision electroweak data. Of course, experiment
has yet to reveal any evidence for SUSY, in
either the sparticle or the Higgs sector.

Unfortunately, supersymmetric models
in general, and the MSSM in particular, provide no explanation for
the number of families (we denote the number
of families/generations by $\ng$) or the patterns of fermion masses. Certainly,
a supersymmetric extension of the standard three-family model is
entirely consistent, but the possibility of incorporating one or more
additional generations into the MSSM framework deserves exploration.
In a previous work,\Ref\GMP{J.F. Gunion, D.W. McKay and H. Pois,
\plbj{334}, (1994) 339.}\ we investigated the
minimal four-family, gauge-unified MSSM, with a large Dirac mass
for the $\nupr$,\Ref\KING{For a detailed discussion of the neutrino sector, see
S. King, \plbj{281}, 295 (1992).}
so as to be consistent
with LEP constraints.\Ref\lepconstraints{OPAL Collaboration, M.Z. Akrawy \etal,
\plbj{236}, 364 (1990); ALEPH Collaboration, D. Decamp \etal, {\ibid},
{\bf 236}, 511 (1990). A recent survey of lower bounds on heavy and
neutral charged leptons and on heavy quarks is given by the
Particle Data Group, K. Hikasa \etal, \prdj{50}, Part 1 (1994).}

In particular, we explored the constraints
upon the model obtained by generalizing
the usual perturbative limit on the top quark
Yukawa coupling $\lambda_t$ to the requirement
that all Yukawa couplings remain perturbative at energy scales below $\mgut$.
Assuming that the CDF and D0 `top-quark' events arise from
a third generation top with $\mt\gsim 155\gev$,\foot{We define the
fourth generation by the CKM
matrix hierarchy $|V_{tb}|^2,|V_{\tpr\bpr}|^2\sim 1$, with $|V_{\tpr
b}|^2,|V_{t\bpr}|^2\ll 1$.}
we demonstrated that perturbative Yukawa behavior requires
a rather light fourth-family spectrum, namely $m_{\tpr,\bpr}\lsim m_t$.
The fourth-family lepton sector is even more strongly constrained:
$m_{\taupr,\nupr}\lsim 85\gev$, the upper limit occurring when $\mtp,\mbp$ are
just beyond the reach of LEP. (That is, the upper
limits for the fourth-family leptons versus quarks
have a strong inverse correlation.)
Additional results included: i) limits requiring $\tanb$
($\tanb\equiv v_2/v_1$, $v_1$ and $v_2$ being the vacuum
expectation values of the $H_1$ and $H_2$ scalar fields)
to have a very modest value, roughly $1\lsim\tanb\lsim 3$,
in order to avoid perturbativity problems for the Yukawas;
ii) the impossibility of imposing
the $\lambda_b(\mgut)/\lambda_\tau(\mgut)=1$ boundary condition in
four-family models; and iii) an increase in the predicted value for $\alphaz$
by $\simeq +3\%$
from the purely gauge coupling contributions to gauge coupling running,
which could, however, be largely compensated by two-loop
Yukawa contributions to gauge coupling running.
\refmark{\GMP}

In this paper we elaborate upon our earlier results and extend our
four family study to include the superpartners of the fourth-family
quarks and leptons.  We adopt the conventional framework in which
the MSSM parameters are determined in the context of minimal supergravity
by universal (at $\mgut$) soft-SUSY-breaking masses for the gauginos ($\mhalf$)
and for the squark and Higgs boson fields ($m_0$),
and universal soft Yukawa couplings ($A$); of course, these typically
evolve to (rather disparate) weak-scale values.
Additional crucial parameters are:
$\tanb$ (defined above) and
$\mu$, the coefficient of the $\hat H_1\hat H_2$ mixing term
in the superpotential. Aside from having
considerable theoretical motivation, this approach has the advantage
that relatively few independent parameters,
$\tanb,\mt,m_0,A,\mhalf $ (along with the sign of $\mu$), are sufficient to
completely specify the theory at the weak scale.
\foot{For a review of this approach, see
\REF\dreesreview{M. Drees and S.P. Martin, `Implications of SUSY
Model Building', Report of Subgroup 2 of the DPF Working Group
on ``Electroweak Symmetry Breaking and Beyond the Standard Model'',
MADPH-95-879.} Ref.~[\dreesreview] and references therein.}

Specific supergravity (SUGRA) and string models make definite predictions
for the relative sizes of $m_0$, $A$ and $\mhalf$.  It is convenient
to specify a given model in terms of the ratios $\xi_A=A/\mhalf$
and $\xi_0=m_0/\mhalf$.  Two models provide particularly
useful benchmarks.  The first is the string-motivated
dilaton boundary condition scenario specified by
$\xi_0=1/\sqrt 3$, $\xi_A=-1$. This boundary condition set emerges universally
in all string models where supersymmetry breaking is dilaton dominated.
It represents a middle-of-the-road choice in that the gaugino mass
$\mhalf$ and the soft scalar mass $m_0$ are both of importance
in the final low-energy values of the squark and slepton masses,
but sleptons are generally significantly lighter than squarks,
and both are generally lighter than the gluino.
A more extreme boundary condition choice is the `no-scale' model
with $\xi_0=\xi_A=0$, which can also arise in certain string
and supergravity approaches.  In this model, supersymmetry breaking arises
entirely from the gaugino mass $\mhalf$ at $\mgut$, with all other
supersymmetry breaking parameters generated by RGE evolution as
the energy scale decreases. Sleptons are still lighter compared to squarks,
and the gluino mass is generally the largest.
A brief review of these two boundary conditions can be
\REF\bgkp{H. Baer, J. Gunion, C. Kao and H. Pois, \prdj{51}, 2159 (1995).}
found in Ref.~[\bgkp].
As a final benchmark possibility we shall also consider $\xi_0=-\xi_A=1$.
In this case, the slepton and squark masses turn out to be large
(often larger than the gluino mass) and
fairly similar in size due to the dominance of the $m_0$ source term.

Our study is designed to complement the existing MSSM and supergravity
(SUGRA) studies, virtually all of which have assumed three generations,
\Ref\mcvetic{An exception is the early $\ng=4$, $\mnup=0$ supergravity
extension of the SM by M. Cvetic and C. Preitschopf, \npbj{272}, 490 (1986).}
and the many earlier studies of a four-family SM.
\Ref\AGRAWALetal{For more recent work, see
P. Agrawal and W. -S. Hou,
\prdj{46}, 1022 (1992); P. Agrawal, S. Ellis and W. -S. Hou, \plbj{256},
289 (1991);  S. King, Ref.~[\KING]; C. Hill and E. Paschos,
{\ibid}, {\bf 241}, 96 (1990); W.-S. Hou and R. G. Stuart, \npbj{349},
91 (1991); M. Sher and Y. Yuan, \plbj{285}, 336 (1992); H. Fritzsch,
\plbj{289}, 92 (1992); B. Mukhopadhyaya and D.P. Roy, \prdj{48}, 2105
(1993); J.L. Hewett, SLAC-PUB-6521 (1994);
Z. Berezhiani and E. Nardi, UM-TH-94-36, hep-ph/9411249 (1994).}
Since the LEP experiments rule out the possibility of an additional
new sequential `light' neutrino,\Ref\Nneutrino{Mark II
Collaboration, G. S. Abrams \etal,
\prlj{63}, 2173 (1989); L3 Collaboration, B. Advera \etal, \plbj{231}, 509
(1989); OPAL Collaboration, I. Decamp \etal, {\ibid}, {\bf 231}, 519
(1989); DELPHI Collaboration, M. Z. Akrawy \etal, {\ibid}, {\bf 231},
539 (1989).}
the fourth-family neutrino must be quite massive, $\mnupr>45\gev$,
and the fourth family would seem to not be truly `sequential'.
However, there are many indications of small non-zero mass
for the neutrinos of the first three families, in which case
the much larger mass for the $\nupr$ is no different than the large
value of the top quark mass as compared to the masses of the other quarks.
Both of these large generational hierarchies must find explanation
in physics beyond the MSSM. A fourth family with a heavy neutrino
is no more unnatural than a third family with a heavy top quark.
\Ref\bnmot{For a recent example of theoretical motivation of $\ng=4$,
see Berezhiani and Nardi, Ref.~[\AGRAWALetal].}

In the minimal supersymmetric model with $\ng \leq4$, the gauge couplings unify
\Ref\unif{P. Langacker, in Proceedings of the
PASCOS-90 Symposium, eds. P. Nath and S. Reucroft (World Scientific, 1990);
P. Langacker and M. -X. Luo, \prdj{44}, 817 (1991); U. Amaldi, W. de Boer, and
H. Furstenau, \plbj{260}, 447 (1991); J. Ellis, S. Kelley and D. V. Nanopoulos,
{\ibid} {\bf 260}, 131 (1991).} perturbatively at a common scale
$\mgut\simeq 2-5\times 10^{16}\gev$. However, as noted earlier,
for $\ng=4$ we must relax the often-imposed
theoretical prejudice that $\lambda_b(\mgut)=\lambda_\tau(\mgut)$
if we are to generate an acceptable prediction for $m_b/m_\tau$
at low energies. We do not regard this as a significant difficulty
since there are many reasons why the Yukawa couplings might {\it not}
be unified at $\mgut$.
In addition to potentially large weak and GUT-scale threshold
effects,\Ref\threshold{P. Langacker and
N. Polonsky, \prdj{47}, 4028 (1993); {\ibid}, {\bf 49}, 1454 (1994);
V. Barger, M. Berger, P. Ohmann, and R. J. N. Phillips,
\plbj{314}, 351 (1993); G. Kane, C. Kolda, L. Roszkowski,
and J. Wells, \prdj{49}, 6173 (1994); G. Ross and R.G. Roberts,
\npbj{377}, 571 (1992); B.D. Wright, MAD/PH/812. For string Yukawa
threshold effects, see I. Antoniadis, E. Gava, K.S. Narain and T.R. Taylor,
\npbj{407}, 706 (1993).}\ a different field content can drastically modify the
GUT-scale relations among the Yukawa couplings. For example, if a
$\bf 45$-Higgs is added to the $SU(5)$ GUT theory, then the strict relation
$\lambda_b(\mgut)/\lambda_\tau(\mgut)=1$ is no longer
valid.\Ref\Frampton{P.H. Frampton, S. Nandi and J.G. Scanio,
\plbj{85}, 225 (1979).}\ More recently, the authors in
\REF\dimop{S. Dimopoulos and A. Pomarol, CERN-TH/95-44, hep-ph/9502397.}
Ref.~[\dimop] have shown that if the theory contains additional heavy
fermions at $\mgut$ then
$\lambda_b(\mgut)/\lambda_\tau(\mgut)<1$ can naturally result.

Thus, it is of considerable interest to study
the phenomenology of a four-family MSSM model.
In our study, we will delineate how current experiments
can either eliminate or confirm the existence of a fourth family.
We shall also discuss a number of theoretical subtleties that arise in
the implementation of four generations in the context of the SUGRA
framework. Aside from updating
the constraints on $\mtpr$, $\mbpr$, and $\tanb$, we will particularly
address the following specific issues.
i) When all basic constraints are applied, what are
the experimentally allowed regions of parameter space?
ii) Does the electroweak (EW) radiative breaking mechanism work in the
$\ng=4$ case?
iii) What are the additional radiative corrections
to the lightest SUSY Higgs mass?
iv) How is the general spectrum of the sparticles affected by the
presence of four families?
v) In particular,
are there new constraints on $\msusy$ arising from fourth-family sparticle
mass constraints?
vi) More generally, how does the fourth family sparticle spectrum
compare to the spectra of the first three families?
And, vii) what experimental constraints
are imposed by the latest $D0$ and CDF top quark searches, as well as a
global fit to the latest EW precision data?
We examine each of
these issues, although not precisely in the above order.

Section 2 is devoted to a study of gauge and Yukawa
coupling unification. In particular, we refine our earlier
analysis to include exact (numerical)
solution of the coupled two-loop gauge and Yukawa couplings.
The parameter space regions given by demanding perturbativity for
the Yukawa couplings and consistency with non-observation
of sparticles and fourth-family fermions at LEP are specified.
Section 3 discusses the radiative electroweak symmetry breaking mechanism.
Section 4 describes the physical Higgs boson masses and
their phenomenology in the four-generation MSSM.
Section 5 discusses a constraint on the SUSY sparticle mass scales
that is peculiar to the four-generation MSSM model.
Section 6 presents, for a number of typical models, the sparticle mass spectra
that arise from a four-family scenario, and delineates allowed
regions of soft-supersymmetry-breaking
parameter space after all direct experimental constraints have
been imposed. Key model-independent features of the sparticle
spectra are identified. Section 7 reviews
the latest direct collider limits on a fourth family, including the
implications of the latest D0 and CDF data.
\REF\habertop{H.E. Haber, preprint CERN-TH/95-178 (1995).}
There, we demonstrate that a scenario\refmark\habertop\  in which
the top is light, $\mt\sim\mw$, but not observed
because it decays by the mode $t\rta {\wt t_1}\cnone$ (the CDF/D0 events coming
from $\tp\rta b W$ in this scenario),
is not consistent with universal soft-SUSY-breaking
boundary conditions by virtue of the Section 5 constraint
which forces the $\wt t_1$ to be very heavy.
In addition, a global fit to
the latest precision LEP data is presented and the ensuing constraints on a
fourth family are discussed. We also explore implications
of a fourth family for the inclusive-jet and di-jet spectra
of the light quarks following from the dramatic slow-down in the evolution
of $\alpha_s$ once energy scales above $\msusy$ are reached.
Section 8 presents our summary and concluding remarks.
The renormalization group formulae and beta functions that are frequently
referenced in the text appear in Appendix A. The question of the
accuracy with which $\tanb$ can be determined by squark/slepton
mass measurements is discussed in Appendix B.

\vskip .5in

\leftline{\bf 2. Gauge and Yukawa Coupling Unification for $\bf \ng=4$}

Gauge coupling unification and the low energy
prediction for $m_b/m_\tau$ (assuming $\lambda_b(\mgut)=\lambda_\tau(\mgut)$)
at the two-loop level in an MSSM four-family model were first considered in
\REF\Nanojones{D. V. Nanopoulos and D. Ross, \plbj{118}, 99 (1982);
J. Bjorkman and D. Jones, \npbj{259}, 533 (1985).}
Ref.~[\Nanojones].
For complete family representations, gauge coupling unification implies
values for $\alphaz$ and $\mgut$ that are independent of
the number of families at the one-loop level. At two-loops,
there is a weak dependence on $\ng$.
\FIG\unification{Gauge coupling unification, including two-loop
gauge contributions but not two-loop Yukawa contributions;
dashed (solid) curves correspond to $\ng=3~(4)$.
The illustration is for $\sinsq=0.2316$ and $\alpha_{em}^{-1}(\mz)=127.9$.}
Fig.~\unification\ shows the unification of the couplings in both
the three- and four-family scenarios including two-loop gauge contributions
to the the beta functions but not including the two-loop Yukawa
contributions.
Assuming the published value of $\alpha_{em}^{-1}(\mz)=127.9$ (in the $\msbar$
scheme), and adopting $\sinsq(\mz)=0.2316$ (see the discussion
of the next paragraph),
\REF\langacker{P. Langacker and N. Polonsky, UPR-0642T}
and a single SUSY breaking scale $\msusy=\mz$,
\foot{As described in Ref.~[\langacker], the effective $\msusy$
is generally not far from $\mz$ even when some superpartners are heavy.}
we predict $\alphazo=(0.1283,0.1326)$, $\alpha^0_U=(0.0432,0.0917)$ and
$\mgut^0=(2.98,5.73)\times 10^{16} \gev$ for $\ng=(3,4)$, respectively.
Here, the superscript `0' indicates that the two-loop Yukawa effects
have not been included. We see that there is a
slight shift in $\alphaz$, a significant shift in
$\mgut$, and a factor of two increase in $\alpha^0_U$
as one moves from $\ng=3$ to $\ng=4$.
For the more recent value\Ref\Swartz{M. L. Swartz, SLAC Report
No. PUB-6710}\ of $\alpha^{-1}_{em}(\mz)=129.08$
($=128.05$ in the $\msbar $ scheme), and adjusting $\sinsq(\mz)$
downwards by $-0.0002$,\refmark\langacker\ to
$\sinsq(\mz)=0.2314$, the resulting predictions
are $\alphazo=(0.129,0.133)$, $\alpha^0_U=(0.0432,0.092)$ and
$\mgut^0=(3.25,6.28)\times
10^{16} \gev$ for $\ng=(3,4)$. In either case, inclusion of
a fourth family raises $\alphaz$ by about $3\%$ when calculated at two-loop
order without including Yukawa contributions to the two-loop beta functions.
We shall shortly return to this issue.

In all the calculations that follow, we shall employ the published value for
the electromagnetic coupling of $\alpha^{-1}_{em}(\mz)=127.9$
in the $\msbar$ scheme and keep the $\msbar$
value of $\sinsq(\mz)$ fixed at $\sinsq(\mz)=0.2316$.
Holding $\sinsq(mz)$ fixed is an approximation.
The actual `best-fit' value of $\sinsq(\mz)$ for a given set of experimental
data depends upon the top-quark mass, the masses of the fourth-family fermions,
the mass of the light SM-like Higgs boson of the model,
and the masses of the superpartners of all four families
\REF\bagger{J. Bagger, V. Matchev and D. Pierce, \plbj{348} 443 (1995).}
(see, for example, Refs.~[\langacker,\bagger]).
If exact coupling constant unification is demanded, and definite
boundary conditions for the soft-supersymmetry-breaking parameters
are specified at the unification scale, the appropriate `best-fit'
value for $\sinsq(\mz)$ value within the four-generation model could
be obtained by a self-consistent iterative procedure
such as described in Refs.~[\langacker,\bagger]. We have chosen
to avoid this complexity for this first study of the four-generation
supersymmetric model.
We also note that we ignore the differences
between $\overline {DR}$ and $\overline {MS}$ couplings and masses.
In particular, although it is the $\overline {DR}$ couplings that are most
naturally required to unify in supersymmetric models, the differences
between $\overline {DR}$ and $\overline {MS}$ couplings are not
significant at scale $\msusy\sim\mz$ compared to other uncertainties.
As described later, we allow a certain
level of `error' in the unification of the coupling constants
in order to account for the remaining experimental uncertainty in
$\sinsq(\mz)$, the slight $\overline {DR}$ {\it vs.} $\overline {MS}$ mismatch
and the (small) variation in $\sinsq(\mz)$ that would
occur as we change 4th-family masses and SUSY parameters.
We will see that the allowed parameter space is only slightly sensitive
to a relaxation of exact unification, and certainly the general
phenomenological features and issues that we discuss would not
be significantly altered by a more precise treatment.

\FIG\allowedregions{We give the allowed $\mtp,\mbp$ parameter
space regions for $\mt(\mt)=165\gev$ in the cases $\tanb=1.5$ and 2.2.
We have taken $\mnupr=\mtaupr=50\gev$.
Full two-loop contributions to the gauge coupling beta functions are included.
Small dots indicate regions of $\mtp,\mbp$ disallowed
by demanding perturbative Yukawas ($\lambda_i\leq 3.3$)
and gauge coupling unification to within 2.5\%.
Small squares indicate additional points excluded if gauge unification
is required to better than 0.01\%.}
\FIG\unificationyuk{Contours in $\mbpr$ and $\mtpr$
(running mass) parameter space of constant $\mgut$ and $\alphaz$
after the inclusion of Yukawa terms in the two-loop gauge coupling
beta functions, assuming $\mt(\mt)=165\gev$ and $\tanb=1.5$.
We have taken $\mnup=\mtaup=50\gev$.}
\FIG\yukawacontours{We display
contours of constant $\lambp$, $\lamtp$ and $\lamtaup$ at $\mgut$
in the region of $\mtp,\mbp$ parameter choices allowed
for 2.5\% unification accuracy with
$\mt(\mt)=165\gev$ and $\tanb=1.5$ or 2.2.
We have taken $\mnupr=\mtaupr=50\gev$.
Two-loop contributions to the gauge coupling beta functions are included.}
The $\mbp,\mtp$ parameter space for $\mnup=\mtaup=50\gev$
(consistent with LEP limits of $45\gev$\refmark\lepconstraints)
and $\mt=165\gev$, equivalent to $\mt(pole)=175\gev$,
\foot{Unless explicitly indicated, up until Section~7
all masses are the running $m(m)$ masses, and not the pole masses.}
that is allowed after demanding perturbativity for the Yukawa couplings for all
energy scales up to $\mgut$, as well as coupling constant
unification at $\mgut$, including two-loop Yukawa
contributions to the gauge coupling evolution equations,
is illustrated for $\tanb=1.5$ and $\tanb=2.2$ in Fig.~\allowedregions.
Our precise criterion for perturbativity is that all Yukawa
couplings obey $\lambda_i\leq 3.3$;
this value ensures that one-loop effects dominate over two-loop
effects.\Ref\BARGERp{See, for example,
V. Barger, M. S. Berger and P. Ohmann, \prdj{47}, 1093 (1993).}\
Two possible levels of gauge unification are considered: unification
of $\alpha_3$ with $\alpha_1$ and $\alpha_2$ to within 2.5\%
and to within 0.01\%. Fig.~\allowedregions\ shows
the allowed parameter space regions for these two cases.
We observe that some of the excluded points on the border become allowed
if the precision demanded for unification of the couplings
is relaxed by even a few percent. Our procedure and these results
will be explained in more detail shortly.

The allowed parameter space regions illustrated in Fig.~\allowedregions\
are even more restricted than those given in our
earlier work, Ref.~[\GMP], at these same $\tanb$ values,
for two reasons: i) the higher value
of $\mt(\mt)=165\gev$ ({\it vs.} $\mt(\mt)=160\gev$)
has been chosen so as to yield $\mt(pole)\sim 175\gev$,
consistent with the latest CDF and D0 experimental
results\Ref\newtopresults{
F. Abe \etal\ (CDF Collaboration), \prlj{74}, 2626 (1995);
S. Abachi \etal\ (D0 Collaboration), \prlj{74}, 2632 (1995.}\
(the precise pole mass value depends
on the sparticle spectrum, but only weakly);
ii) the two-loop Yukawa effects in the running of the gauge couplings
feed back into the Yukawa couplings themselves
so that they violate perturbativity more easily.
This latter point can be
understood by noting that inclusion of Yukawa couplings
reduces the value of $\alpha_3(Q)$ when $\mtp,\mbp$ have values
near the boundary of the perturbatively allowed region.
The reduction in $\alpha_3(Q)$ in turn slightly reduces the magnitude of the
negative gauge contribution to the one-loop component
of the Yukawa beta functions as $Q$ runs from $\mz$ to $\mgut$
(see Eqs.~(A.1), (A.5) and (A.6) in Appendix A),
constraining the weak-scale value of the Yukawas to be somewhat smaller
than before in order to remain perturbative at $\mgut$.

This reduction in $\alpha_3$ for points at the parameter
space edge is illustrated in
Fig.~\unificationyuk\ which shows how much the values of $\mgut$ and $\alphaz$
can be affected by including the {\it large} Yukawa contributions to the
two-loop gauge coupling beta functions in the four-family scenario.
The values of $\mgut$ and $\alphaz$ are presented as contours in the
the $m_{\tpr},m_{\bpr}$ plane.
\foot{Not shown is $\alpha_U$; the shift in $\alpha_U$ due to
inclusion of two-loop Yukawa contributions is
comparable to the $\alphaz$ shift, {\it i.e.} $\lsim 4\%$.}
The shifts in $\alphaz,\mgut$ from $\alphazo,\mgut^0$ agree well with our
approximate solutions from Ref.~[\GMP].
For $\mtp,\mbp$ near their maximal allowed
values, $\alphaz,\mgut$ can be lowered by as much as $5\%,16\%$ respectively.
We see that the $\ng=4$ values of $\alphaz$ are essentially the
same as the $\ng=3$ value when large Yukawa effects are included at two-loop.

The precise boundary of the perturbatively allowed region depends
upon the precision demanded for coupling constant unification.
In Fig.~\allowedregions\ we showed the allowed regions obtained by
demanding unification of the coupling constants to within 2.5\%
or $0.01\%$ (the latter being essentially equivalent to exact unification).
The allowed region in the $0.01\%$ case is slightly reduced compared to the
$2.5\%$ case. This sensitivity  is
implicitly present even before the two-loop Yukawa terms are included in
gauge coupling evolution.  As noted above, by increasing $\alpha_3$
the Yukawa blow-up is delayed.  So if $\alpha_3(\mgut)$ is allowed
to be slightly larger than the common value of $\alpha_1(\mgut)$
and $\alpha_2(\mgut)$ (denoted $\alpha_{1,2}(\mgut)$), then slightly
larger values of $\mtp,\mbp$ will be allowed by the requirement
of perturbativity for Yukawa couplings up to $\mgut$.
When two-loop Yukawa couplings are included in the gauge evolution equations,
an iterative procedure must be employed for finding a fully consistent
solution. For a given $\mtp,\mbp$ choice and a given possible starting
value of $\alpha_3(\mz)$, full two-loop evolution of all three
couplings may be performed (which requires evolving also the Yukawas
at the same time).
The value of $\mgut$ at which $\alpha_1$ and $\alpha_2$
unify can then be determined. The common value, $\alpha_{1,2}(\mgut)$, can
then be compared to the evolved $\alpha_3(\mgut)$ value.
This process is iterated until the evolved $\alpha_3(\mgut)$ value
is as close to $\alpha_{1,2}(\mgut)$ as possible without the
Yukawas becoming too large ($\lambda_i<3.3$ is required).
The precision of unification is then specified by the requirement
that $\alpha_3(\mgut)=\alpha_{1,2}(\mgut)$ to within a definite percentage
deviation: for example unification to within 2.5\% means that
$\alpha_3(\mgut)$ is allowed to be no more than 2.5\% larger than
$\alpha_{1,2}(\mgut)$. Given experimental errors
in $\sinsq(\mz)$ and variations
in its best fit value as fourth-generation fermion masses and sparticle masses
for all the generations are varied, we regard the region allowed by
unification to within 2.5\% as fully acceptable. However, to go much beyond
the 2.5\% allowed region of Fig.~\allowedregions\ would almost certainly
require accepting the fact that one of the Yukawa couplings
becomes non-perturbative at a scale below $\mgut$ or
that $\alpha_3(\mgut)$ truly exceeds $\alpha_{1,2}(\mgut)$, \eg\
due to non-renormalizable operators,
string threshold effects and/or evolution between $\mgut$ and
$M_{string}$.

Of course, as at one loop, the value of $\alphaz$ consistent with
unification is sensitive to the effective
scale $\msusy$ implied by the sparticle masses.
As the value for $\msusy$ is raised, $\alphaz$ is reduced; for $\ng=3$ and
$\msusy=1\tev$, the shift can be as large as $\simeq -10\%$.
Similar results hold for $\ng=4$ as well. Given the present measurement of
$\alphaz=0.12\pm 0.01$, it would seem
that both the $\ng=3$ and $\ng=4$ scenarios
predict an $\alphaz$ which is somewhat high unless $\msusy$ is
significantly above $\mz$. However, a large value of the effective
$\msusy$ is not easy to achieve,\refmark\langacker\
when the slepton, neutralino and chargino mass scales are significantly
lower than the squark and gluino mass scales.
Nonetheless, it is fair to say that
$\alphaz$ per se does not discriminate between the $\ng=3$ and $\ng=4$ cases,
since the increase in $\alphaz$ for $\ng=4$ {\it vs.} 3
from pure gauge effects can be compensated in the $\ng=4$ case by
the large Yukawa effects in the gauge running, as just discussed.
In addition, we shall find that $\ng=4$ scenarios
characteristically force {\it all} the sleptons
and squarks to higher masses (including those of the first three families),
implying that the effective $\msusy$ could
be larger for $\ng=4$ than for $\ng=3$. Given this and the
additional uncertainties associated with weak and GUT-scale threshold effects,
we believe that the $\ng=3$ and $\ng=4$ scenarios are equally admissible.
Attempts to address the moderately high prediction for $\alphaz$
(that emerges in both cases) have recently been
considered in Refs.~[\langacker,\bagger].

In general, it is important to note that two-loop contributions
to the running of $\alpha_s(Q)$ will be much more significant
in the case of $\ng=4$ than for $\ng=3$.
This is simply due to the fact that the one-loop beta function for $\alpha_s$
is proportional to $(9-2\ng)$ (once $Q>\msusy$) and therefore is
rather small for $\ng=4$. This means that two-loop
contributions can represent a much larger percentage of the total
beta function than in the $\ng=3$ case.  The importance
of two-loop contributions will be especially apparent when
considering the running of the squark masses, where some
terms involve the running value of $\alpha_s^3$.  The cumulative
effect of two-loop contributions can be large in such cases
when evolving all the way down from $Q=\mgut$ to $Q=\mz$.
The situation for $\alpha_1$ and $\alpha_2$ is quite the opposite.
Indeed, the one-loop beta function for $\alpha_2$ {\it increases}
by a factor of 3 in going from $\ng=3$ to $\ng=4$, implying that
$\alpha_2$ increases much more rapidly as the energy scale is
varied from $\mz$ up to $\mgut$.

What about Yukawa coupling constant unification? In Fig.~\yukawacontours,
contours of constant $\lambp(\mgut)$, $\lamtp(\mgut)$ and $\lamtaup(\mgut)$
within the allowed parameter space region are shown.
{}From these contours we discover a number of important facts.
First, $\lambp(\mgut)$ and $\lamtaup(\mgut)$ are generally quite different;
Yukawa unification does not generally occur, although
we see that in the $\mtp,\mbp\sim 100\gev$ corner of the $\tanb=1.5$
plot we do have $\lamtp\sim\lambp\sim\lamtaup$.
Second, we see that most of the non-perturbative borders are
defined by one of the fourth-family Yukawa couplings becoming
non-perturbative. At $\tanb=1.5$ the right-hand border
results from non-perturbative behavior for $\lambp$
(at the upper boundary, $\lamt$ becomes non-perturbative)
while at $\tanb=2.2$ the right-hand border results when
$\lamtaup$ becomes large and the upper boundary arises from
non-perturbative behavior of $\lamtp$.  However, it is also clear
that, in general,
 not all of the fourth-family Yukawas (or $\lamt$) are simultaneously
large.  This means that one is unlikely to be particularly close
to a Yukawa fixed-point solution\Ref\fixedpoint{B. Pendelton
and G. Ross, \plbj{98}, 291 (1981); C. Hill, \prdj{24}, 691 (1981);
W. Bardeen, M. Carena, S. Pokorski and C. Wagner,
\plbj{320}, 110 (1994); J. Bagger, S. Dimopoulos and E. Masso,
\prlj{55}, 920 (1985).}\ such that the low-energy values of
the $\lam$'s are rather independent of their $\mgut$-scale values.
The nearest approach to a fixed point occurs in the
$\mtp\sim\mbp\sim 100\gev$ corner of the $\tanb=1.5$
figure; the actual fixed point location is indicated by an $X$.
In our earlier work,\refmark{\GMP} with $\mt(\mt)=160\gev$
and two-loop Yukawa contributions to gauge running not included
in determining the parameter space,
the large $\mtp,\mbp$ corner of the allowed region was nearer to
this fixed point.

\bigskip

\leftline{\bf 3. EW Symmetry Breaking}

One of the many nice features of the MSSM extended by minimal SUGRA is
that, for a very large region of soft-SUSY-breaking parameter space,
radiative EW-symmetry breaking is automatically
induced by renormalization group evolution. Thus, the hierarchy
between $\mgut$ (where scalar masses are universal and (therefore) EW symmetry
is initially unbroken) and $\mz$ receives a natural
explanation. In the three-family model, this EW-symmetry
breaking is mainly a result of the quantum
corrections arising from the large $\lamt$
Yukawa coupling which drives the $H_2$-Higgs field squared-mass
to ever smaller values
as the energy scale is decreased; EWSB occurs when the Higgs mass-squared term
in the scalar Higgs potential is finally driven to a negative value.
Of course, to obtain the precise value of $\mz$ as given by
$\mz^2={1\over 2}(g^2+g'^2)(v^2_1+v^2_2)$,
there must be a relation
among the supersymmetric model parameters. Mathematically,
these relations result from requiring that the first derivatives
of the scalar field potential with respect to $v_1$ and $v_2$ vanish.
Normally, one of these conditions is used to determine the magnitude
(but not the sign) of $\mu$ in terms of the other initial parameters of
the theory, including the soft-SUSY-breaking parameters.
The other condition is used to determine the magnitude of $m_3^2$,
appearing in the
$-m_3^2(H_1H_2+{\rm h.c.})$ mixing term in the scalar field potential.
Thus, $\mu$ and $m_3^2$ at the weak scale become functions of $\mtp$, $\mbp$,
$\tanb$, $\mz$, $m_0$, $\mhalf$ and $A$.
Since this radiative breaking mechanism is {\it
essential} to the viability of the MSSM plus minimal SUGRA, we describe
its dynamics for $\ng=4$ before turning to predictions
for the SUSY sparticle spectra, and experimental
constraints from direct particle searches and EW precision measurements.
Not surprisingly, we find that in a four-family scenario the additional large
Yukawa couplings also feed into the running of the Higgs field squared-masses,
and modify the running significantly.
To indicate the way in which radiative breaking occurs in the
four-family case, it is convenient to present the discussion
at tree-level.  A tree-level
discussion is adequate for general understanding,
and is a good approximation so long as the tree-level minima
obey all of the necessary stability and consistency constraints.
\Ref\grz{See, for example, G. Gamberini, G. Ridolfi and F. Zwirner,
\npbj{331} 331 (1990).}
However, we emphasize that
our full numerical calculations are actually performed by minimizing
the full scalar potential at one-loop.

\FIG\radiativebreaking{In a) we illustrate
the evolution of the soft-SUSY-breaking scalar mass-squared
parameters $m_{H_1}^2$ and $m_{H_2}^2$ as well as the value of $|\mu|$.
In b) we show the evolution of the coefficients $m_1^2$, $m_2^2$ and
$m_3^2$ of the $H_1^2$, $H_2^2$ and $2H_1H_2$ scalar field potential terms,
as well as that of $S\equiv m_1^2m_2^2-m_3^4$. We have taken $\xi_0=1/\sqrt3$,
$\xi_A=-1$, $\mhalf=400\gev$, $\tanb=1.5$,
$\mtp=100\gev$, $\mbp=100\gev$, and $\mtaup=\mnup=50\gev$. All masses
are given in units of $\mhalf$.}
Figures~\radiativebreaking a and \radiativebreaking b illustrate
how radiative breaking occurs
in the tree-level approximation in one specific four-family case.
We take $\mt=165\gev$,
$\tanb=1.5$, $\mtp=100\gev$, $\mbp=100\gev$ and $\mtaup=\mnup=50\gev$.
As outlined in the introduction,
we specify our soft-SUSY-breaking parameter boundary conditions in terms of
$\xi_0\equiv m_0/\mhalf $, $\xi_A=A/\mhalf$ and $\mhalf$.
For this illustration we have adopted the string-motivated
dilaton boundary condition scenario specified by $\xi_0=1/\sqrt 3$, $\xi_A=-1$.
In the plots, $m_{H_1}^2,m_{H_2}^2$ are the Higgs mass-squared parameters
appearing in the soft-SUSY-breaking potential
(at $\mgut$ they are equal to $m_0^2$), while $m_{1,2}^2=
m_{H_{1,2}}^2+\mu^2$ are the parameters multiplying the $H_1^2$ and $H_2^2$
quadratic scalar field terms in the {\it full} scalar field potential;
as defined earlier, $m_3^2$
is the coefficient of the $H_1H_2$ mixing term in the scalar field potential.
In Fig.~\radiativebreaking, we see that the
$m^2_{H_i}$ start out above zero at $Q=\mgut$, and then
evolve below $\mgut$ so that {\it both} eventually
take on negative values (but with $m^2_{H_2}$
being the more negative). The $\mu$ parameter is determined
at the weak scale by minimization of the RGE-improved tree-level
Higgs potential, and is then evolved up to $\mgut$. The fact that
$\mu^2(\mz)>0$ for the chosen value of $\tanb$ and the given value of $\mz$
indicates that an allowed
four-family EW symmetry breaking solution via the radiative breaking
mechanism exists.

However, some differences in comparison to $\ng=3$ are apparent. Even
for the small $\tanb=1.5$ value,
$m_{H_1}^2$ has significant evolution due to the large
$\lambp$ and $\lamtaup$ Yukawa couplings. In the standard three-family case,
$m_{H_1}^2$ only evolves significantly
when $\tanb$ is so large as to require a large value for $\lamb$.
In Fig.~\radiativebreaking b, one can see that although the scalar potential
parameters $m_1^2$, $m_2^2$ and $m_3^2$
each evolves separately and ends with a value that is
$>0$, in combination they serve to trigger the breaking of the EW symmetry,
measured by the tree-level stability condition ${\cal S}=m^2_1 m^2_2-m^4_3$.
It is of course possible for the mechanism to fail, i.e. if $\mu^2<0$.
This can happen in the case where $m^2_{H_1}$ is too large compared to
$m_{H_2}^2$.

\bigskip

\leftline{\bf 4. The $\bf \ng=4$ Higgs Sector}

We begin\foot{An excellent brief overview
of Higgs phenomenology and discovery techniques
is now available in
\REF\dpfshort{T.L. Barklow \etal, `Electroweak Symmetry Breaking
and Beyond the Standard Model', SLAC-PUB-95-6893 (1995).}
\REF\dpfhiggs{J.F. Gunion, A. Stange and S. Willenbrock, in preparation.}
the Higgs subgroup summary appearing in Ref.~[\dpfshort]. A longer
version of this review will soon be available, Ref.~[\dpfhiggs].
References for statements
not explicitly referenced below can be found in these reports.}
by reminding the reader that at tree-level the Higgs
sector is determined by just two parameters, $\tanb$ and $\mha$
(the mass of the CP-odd scalar Higgs boson), and that if $\mha$
is large (as we shall see it is in all the SUGRA models considered)
then the lightest CP-even Higgs boson $\hl$ is very SM-like
and has mass bounded from above by $\mz$.  However, it is well-known
that one-loop radiative corrections to $\mhl$ can significantly increase
the upper limit of $\mhl$ (denoted $\mhlmax$) for the large $\mt$ value found
in the CDF and D0 experiments. Exactly how large the upper bound
is depends upon other SUSY parameters, the most important sensitivity
being to the stop squark mass, $\mstop$. In the absence of a fourth
generation, if the running masses are
$\mstop=1\tev$ and $\mt=165\gev$, then $\mhl$
can be as large as $120-125\gev$ at large $\tanb$. However,
in many SUGRA models (for example the dilaton scenario) $m_0$
and, consequently, $\mstop$ are significantly smaller than $1\tev$
for $\mgl$ values below 1 TeV.  In such models $\mhlmax$
is typically $\lsim 100\gev$.

The magnitude of $\mha$ determines the observability
of the $\ha$ and $\hh$ Higgs bosons, which are more or less degenerate
when $\mha$ is large.  For example, $\epem$ colliders can only
probe up to $\mha\sim\mhh\sim \sqrt s/2-30\gev$
in the $Z^*\rta \hh\ha$ mode (the only viable mode when
$\hl$ is SM-like). For $\ng=3$,
SUGRA model predictions for $\mha$ can range from $200\gev$ on up,
and an $\epem$ collider with $\sqrt s=500\gev$ would at least have
a small chance of seeing $\hh+\ha$ production.
The parameter $\mu$ is also of interest in that its origin
in the SUGRA models is rather uncertain.  For $\ng=3$, $\mu$
tends to take on rather moderate values $\sim 500\gev$.\refmark\langacker\
We now discuss the influence of a fourth generation.

First, we note that a fourth family will give an additional set of
$\taup,\staup$, $\nup,\snup$, $\bp,\sbp$
and $\tp,\stp$ loop contributions to the radiative corrections
for $\mhlmax$.  (As noted earlier, as a consequence of the many loop
corrections for $\ng=4$ it is essential that the
one-loop effective potential be used in the scalar
potential minimization as well as in
the determination of the Higgs masses.)
A fourth family will also typically lead to a rather
high $\mgl$ scale, as discussed in the following sections;
this will in turn influence $\mha$ and $\mu$, which will often
take on relatively large values.

\FIG\higgsnewlimit{Contours of $\mu$, $\mha$ and $\mhl$ in
the $\mtp,\mbp$ parameter plane for $\tanb=1.5$, $\mt(\mt)=165\gev$
and $\mhalf=600\gev$.  We give results for
the $\mgut$-scale dilaton boundary conditions: $\xi_0=1/\sqrt3$, $\xi_A=-1$.
The diamonds indicate regions of parameter space disallowed because
$\mstaupone<{\rm max}\left\{45\gev,\mcnone\right\}$ at energy scale $\mz$.}

Typical results for $\mu$, $\mha$ and $\mhl$
are illustrated in Fig.~\higgsnewlimit, where we
display contours of constant
$\mu$, $\mha$ and $\mhl$ in $\mtp,\mbp$ parameter space for
$\mt(\mt)=165\gev$, $\tanb=1.5$ and $\mhalf=600\gev$
in the dilaton scenario: $\xi_0=1/\sqrt 3$, $\xi_A=-1$.
We see that $\mu$
is generally above $1\tev$, a substantial increase over the corresponding
$\ng=3$ dilaton scenario result; $\mha$
is typically also rather large, ranging from $\sim 600\gev$ to
above $1\tev$ --- a $\sqrt s=500\gev$ $\epem$ collider would not allow
detection of the $Z^*\rta \hh\ha$ pair production process.
The large values of $\mha\sim\mhh$ imply that one must also be
cautious to account for supersymmetric decays of the $\ha$ and $\hh$.
These decays can deplete the more easily observed $b\anti b$ decay channel,
especially given that $\tanb$ cannot be large for $\ng=4$,
implying that the $\hh,\ha\rta b\anti b$ coupling cannot be greatly
enhanced relative to the SM-like result. We will not attempt
a detailed study of the $\hh$ and $\ha$ decays here.

The above results are not greatly altered by changing the $\mgut$
boundary conditions, keeping $\mhalf$ fixed at $600\gev$.
In the high-$m_0$ ($\xi_0=1$) scenario, both $\mha$ and $\mu$ become
somewhat ($\sim 100\gev$)
larger over most of the $\mtp,\mbp$ parameter space.  The only exception
is the corner where $\mbp$ is big and $\mtp$ is small; in this corner
$\mha$ is slightly smaller ($564\gev$ compared to $610\gev$)
in the high-$m_0$ scenario than in the dilaton scenario.
In the case of the no-scale boundary conditions, $\xi_0=\xi_A=0$,
both $\mu$ and $\mha$ move to lower values (shifts are of order $50-100\gev$).
For example, in the small $\mtp$, large $\mbp$ corner
one finds $\mha\sim 450\gev$.

Fortunately as regards the prospects for $\hl$ detection,
the extra loops from the fourth family do not
yield overwhelmingly large radiative corrections to $\mhlmax$.
They increase $\mhlmax$ by 10 to 25 GeV relative to corresponding
$\ng=3$ predictions. As illustrated by the contours
in Fig.~\higgsnewlimit, at worst $\mhlmax\sim 130\gev$ at the
perturbative boundary in the $\mbp,\mtp$ parameter space.
For all the three $\mgut$ scenarios, $\mhl$ remains
very much in the $115-122\gev$ range for most of allowed $\mbp,\mtp$
parameter space, for this relatively large $\mhalf=600\gev$ value.
(Lower $\mhl$ values are predicted at lower $\mhalf$.)
The phenomenology of the $\hl$ depends upon whether or not it is SM-like.
For the bulk of parameter space, and certainly for the preferred
scenarios and portions of parameter space, $\mha$ is large and
the $\hl$ will be very SM-like. We discuss its phenomenology
assuming that this is the case.

Because $\mhl$ is predicted to be $\gsim 100\gev$ for many $\ng=4$
scenarios, LEP II would be less likely to find the $\hl$
if there is a fourth family.  However, a $\sqrt s=500\gev$ $\epem$ collider
would have no difficulty in doing so in the $Z^*\rta Z\hl$
production mode.
At the TeV$^\star$ upgrade of the Tevatron,
detection of a SM-like $\hl$ is probably only possible
in the $W\hl\rta \ell\nu b\anti b$ mode, and then only if $\mhl\lsim 95\gev$.
(Although it is not impossible that the $W\hl\rta \ell\nu\tau^+\tau^-$
mode could be used for $110\lsim \mhl\lsim 120\gev$.\Ref\mrenna{S. Mrenna
and G. Kane, CALT-68-1938 (1994).})
For the many four-family scenarios that lead to $\mhl$ above $100\gev$,
searches for the $\hl$ at the Tevatron would be, at best, problematical.

At the LHC, a SM-like Higgs boson (for which we use
the generic notation $\h$ below --- the $\hl$
might or might not be perfectly SM-like) in the mass region being discussed
would typically be found through production via $gg\rta \h$
and decay to either $\gam\gam$ or $ZZ^*$ (with $ZZ^*\rta 4\ell$).
The $gg\rta\h$ production rate, proportional to $\Gamma(\h\rta g g)$
would be greatly enhanced by the additional
$\tp$ and $\bp$ loop contributions to the one-loop $gg\rta\h$ coupling.
\FIG\hdiscovery{We plot the ratio of $\ng=4$ to $\ng=3$ values
for: $\Gamma(\h\rta gg)$, $\Gamma(\h\rta \gamma\gamma)$,
and $\Gamma(\h\rta gg)\times BR(\h\rta \gamma\gamma)$.  We adopt
the scenario of $\tanb=1.5$, $\mt=165\gev$, $\mtp=\mbp=100\gev$,
$\mtaup=\mnup=50\gev$, and assume superpartners are sufficiently
heavy that their contributions to these one-loop quantities are small.}
This is illustrated in Fig.~\hdiscovery\ where we plot for a SM-like $\h$
the ratio of $\ng=4$ to $\ng=3$ values
for: $\Gamma(\h\rta gg)$, $\Gamma(\h\rta \gamma\gamma)$,
and $\Gamma(\h\rta gg)\times BR(\h\rta \gamma\gamma)$, taking
$\tanb=1.5$, $\mt=165\gev$, $\mtp=\mbp=100\gev$,
$\mtaup=\mnup=50\gev$, and assuming that superpartners are sufficiently
heavy that their contributions to these one-loop quantities are small.
(As discussed in Sec.~7a, experimental limits tend to prefer this
type of scenario.) For $\Gamma(\h\rta gg)$, and hence
the $gg\rta \h$ production rate, we see an enhancement by a factor of 10.
Thus, the $4\ell$ channel, which for $\ng=3$ is only viable for
$\mh\gsim 130\gev$, would yield a detectable signal
down to somewhat lower masses, perhaps as low as $\mh=120\gev$. ($BR(\h\rta
ZZ^*)$ falls very rapidly with decreasing $\mh$ so very few events
would result for $\mh$ values much below this.)

The $\h\rta \gam\gam$ decay also arises at one-loop.  For $\ng=3$,
the main contribution is from the $W$-loop diagram.  Fermion loops
(for massive fermions) cancel against the $W$-loop contribution
and decrease the $\h\rta\gam\gam$ width.
For $\ng=4$ this cancellation
can be quite substantial, as illustrated in Fig.~\hdiscovery.
\Ref\gungamgam{These effects were first explored in
J.F. Gunion and S. Geer, {\it
Proceedings of the ``Workshop on Physics at Current Accelerators and
the Supercollider''}, eds. J. Hewett, A. White, and D. Zeppenfeld,
Argonne National Laboratory, 2-5 June (1993), ANL-HEP-CP-93-92, p.~335.}\
Combining the resulting reduction in $BR(\h\rta\gam\gam)$
with the enhanced $gg\rta \h$ production
rate, the resulting $\gam\gam$ channel event rate tends to be
substantially suppressed relative to the $\ng=3$ rate
in the $100-130\gev$ mass range of interest.  In fact, the preferred
$\mtp$, $\mbp$, $\mtaup$ mass choices
delineated above are about the worst that can be made
in this regard. Thus, detection of the $\hl$ in the $\gam\gam$
channel at the LHC becomes problematical, even
when $\mha$ is large and the $\hl$ is SM-like.

Not investigated to date is whether the enhanced production rate
from $gg\rta \h$ might make detection of a SM-like $\hl$ in the inclusive
$b\anti b$ channel possible (assuming high $b$-tagging efficiency and purity).
Finally, we note that
the extra family has essentially no impact on the $pp\rta t\anti t\hl\rta \ell
b\anti b b\anti b X$ LHC detection mode, which would continue to be viable
for $\mhl\lsim 120\gev$, when the $\hl$ is SM-like.\refmark\dpfshort\

Regarding the heavier $\ha$ and $\hh$, for our typical scenario
they are so massive that perhaps the only accelerator with adequate
energy for their production and possible detection
will be the LHC. At the LHC, for $\ng=3$ the
detection of a massive $\ha$ or $\hh$ is possible only if $\tanb$
is so large ({\it e.g.} $\gsim 10$), that the $gg\rta b\anti b\hh$
and $gg\rta b\anti b \ha$ production rates significantly exceed the
inclusive $gg\rta \hh,\ha$ rates.\refmark{\dpfshort,\dpfhiggs}\
However, for $\ng=4$
the $\bpr$ and $\tpr$ loop contributions to $gg\rta\ha$ and
$gg\rta\hh$ will greatly increase these inclusive production rates, regardless
of the $\tanb$ value. The possibility of observing
the $\hh$ and $\ha$ at the LHC in the inclusive $b\anti b$ final state assuming
such highly enhanced rates should be carefully examined.

As a final aside, we note that the standard relation,
\Ref\DN{M. Drees and M.M. Nojiri, \prdj{45}, 2482 (1992).}
$$\mha^2\simeq {{m^2_{\wt \nu}+\mu^2}\over \sin^2\beta},$$ valid for
$\tanb\lsim 20$ if $\ng=3$, is not necessarily maintained
for $\ng=4$ since there will be corrections involving the
$\lambp,\lamtaup,\lamb$ Yukawas (where $\lamb$ can
essentially be neglected). In Fig.~\higgsnewlimit,
for $\mtp,\mbp$ large, {\it e.g.} $\mtp=\mbp=100\gev$, the
relation is satisfied to
within $3\%$; however for $\mbp\simeq 110\gev, \mtp=55\gev$, the relation
is seriously violated: $(\mha^2 \sin^2\beta)/(m^2_{\wt \nu}+\mu^2)\simeq 0.25$.
This latter situation is realized in the region where
$\mbp>\mtp$, a region that is experimentally disfavored unless an unnatural
quark mixing pattern exists (see Ref.~[\GMP] and Sec.~7 for a
more detailed discussion of this point).

\bigskip

\leftline{\bf 5. A Fourth-Generation Sparticle Constraint
on the SUSY Scale}

We turn now to an important additional constraint on soft-SUSY-breaking
parameters that can arise from consistency of the sparticle mass spectrum
with LEP limits and a neutral LSP (lightest supersymmetric particle).
(A charged LSP is excluded experimentally.)
A glance at Eqs.~(A.11) through (A.19) in the Appendix shows that $dm^2_i/dt$
(where $t={1\over 2\pi}\ln[Q(\gev)]$) for squarks and sleptons receives
positive
contributions from Yukawa terms and negative contributions from gauge
terms. Thus, starting from a universal $m_0^2$ and evolving downwards
in $t$ to $\mz$, the lightest squark/slepton will be the one
with the largest Yukawa contributions relative to gauge contributions.
This turns out to always be the $\staup_R$ or the $\snup_R$.  After including
$\staup_R-\staup_L$ mixing, the lightest $\staup$ eigenstate is denoted
$\staupone$, and similarly $\snupone$ is the lightest
$\snup$ eigenstate.

The $\snupone$ can be even lighter than the $\staupone$ in scenarios
with small $\xi_0$, such as the dilaton and no-scale models.  However,
we see no general phenomenological reason for not allowing the $\snupone$
to be the LSP.  Indeed, it is even quite likely that the $\cnone$
decays invisibly when $\mcnone>\msnupone$ via $\cnone\rta \nu_\tau \snupone$,
assuming at least a small non-zero value for the required $3-4$ generation
mixing angle. Thus, we will only impose a phenomenological limit
on the charged $\staupone$.

Eq.~(A.15) shows that the Yukawa contributions
to $d m^2_{\staup_R}/dt$ (in the Appendix we use the notation
$\staup_R={E^\prime}$)
are controlled by ${\cal D}_{\taup}=m_{H_1}^2+m^2_{\staup_L}+m^2_{\staup_R}+
A_{\staup}^2$. For moderate initial values, $m_{\staup_L}=m_{\staup_R}=m_0$,
the more negative $m_{H_1}^2$ is (see Fig.~\radiativebreaking)
the larger will be $m^2_{\staup_R}$; $m_{H_1}^2$ in turn becomes more negative
for larger $\mbp$.  Thus, for some choices of soft-SUSY-breaking parameters
it is possible that $\mstaupone$ will fall below the $45\gev$ LEP limit
and/or below the mass of the lightest supersymmetric particle (the
lightest neutralino, $\cnone$, in the models we consider)
for low values of $\mbp$ (but not for higher values).

To illustrate this, we return to Fig.~\higgsnewlimit, where we have chosen
$\tanb=1.5$, $\mt(\mt)=165\gev$ and dilaton boundary
conditions for the soft-SUSY-breaking parameters with $\mhalf=600\gev$.
The diamonds indicate the portion of parameter space at low $\mbp$
that is ruled out because one predicts $\mstaupone<45\gev$ or
$\mstaupone<\mcnone$.
(For this particular choice of $\mhalf$,
$\mcnone\sim 108\gev$ and it is the $\mstaupone>\mcnone$
requirement that fixes the diamond region.)
By raising $\mhalf$, this problem region is moved to lower $\mbp$
values since the starting value of $m_0^2\propto\mhalf^2$ for
$m^2_{\staup_R}$ is
increased more rapidly than the off-diagonal mixing term $A\propto \mhalf$.
Conversely, by lowering $\mhalf$ we eventually reach a value for
which no portion of parameter space remains allowed.

The type of boundary condition applied is also important. For instance,
for the same $\mhalf=600\gev$ but $\xi_0=1$, $\xi_A=-1$ (the high-$m_0$
scenario) no points with $\mbp,\mtp>50\gev$ are eliminated by
virtue of the $\staupone$ constraint, whereas for the no-scale
choice of $\xi_0=\xi_A=0$, the portion of parameter
space removed expands to include slightly higher values of $\mbp$
than in the dilaton scenario.

Let us further expand upon this point.
{}From the above discussion we see that in the generic one-loop formula,
$m^2_i=m^2_0+\hat C_i \mhalf^2+
D_i\mz^2\cos 2\beta$, the evolution parameter $\hat C_i$ for
$i=\staup_R$ is strongly affected by indirect effects from $\lambp$,
due to the absence of an $\alpha_s$ contribution to $\hat C_i$.
In general, larger $\mbp$ tends to raise $\hat C_{\staup_R}$.
As we have discussed above, this can be traced
to an increasingly negative contribution in the RGE for $m^2_{\staup_R}$
from the increasingly negative value of $m^2_{H_1}$ as $\mbp$ increases.
\FIG\msqevolve{Evolution of $m^2_{\staup_R}$,
$m^2_{\stau_R}$, and $m^2_{\snup_R}$ (in units of $\mhalf^2$).
We have chosen $\tanb=1.5$, $\xi_0=1/\sqrt 3$,
$\xi_A=-1$, and $\mhalf=600\gev$. Results for $\mtp=100\gev$
and $\mbp=90\gev$ are compared to those for $\mtp=100\gev$ and $\mbp=110\gev$.
The corresponding $\mz$-scale $\mstaupone$ values are 189 and $311\gev$,
respectively.}
Fig.~\msqevolve\ demonstrates this sensitivity
of the $\mz$-scale value of $m^2_{\staup_R}$ to $\mbp$
in the case of $\mtp=100\gev$, $\tanb=1.5$, $\mhalf=600\gev$
with dilaton boundary conditions:
as $\mbp$ is lowered from $\mbp=110\gev$ to $\mbp=90\gev$, the
$\mz$-scale value of $m^2_{\staup_R}$ is lowered by
$\sim 40\%$, with the lowest $\staup$ physical eigenstate mass
decreasing dramatically, from $\mstaupone=311\gev$ to $189\gev$.
In contrast, the evolution and $\mz$-scale values of
$m^2_{\stau_R}$ and $m^2_{\snup_R}$ are little affected.
Note the subtlety of the ${\rm log}(Q)$ behavior of $m^2_{\staup_R}$. In
the region of small (and decreasing) ${\rm log}(Q)$
$m^2_{\staup_R}$ {\it rises} due to the increasingly more negative value
for $m^2_{H_1}$ as $\log(Q)$ decreases (see Fig.~\radiativebreaking).
Since this rise is less for smaller $\mbp$,
for small enough $\mbpr$ one can even obtain $m^2_{\staup_R}<0$, and
EM will no longer be unbroken. However, as $\mbpr$ is decreased
one first arrives at a point where
either  $\mstaupone <\mcnone$ and the LSP is no longer neutral,
or $\mstaupone<45\gev$, violating LEP limits.
{\it We re-emphasize that these
requirements result in the strongest additional
RGE-related phenomenological constraint deriving from the SUGRA
extension of the four-family model.}

\FIG\limit{We display the $\staupone$ constraint boundaries
for a variety of $\mgut$-scale scenarios. Regions to the left
of the boundaries are disallowed. We have taken
$\mt=165\gev$, $\mtaup=\mnup=50\gev$, and $\tanb=1.5$.}
To gain some additional insight regarding the stringency of this bound
as a function of $\mgut$-scale boundary conditions,
we present the right-hand boundaries, analogous to that of
the diamond region shown in Fig.~\higgsnewlimit,
for a selection of different
possibilities in Fig.~\limit. The right-hand window shows
results for the dilaton case of $(\xi_0,\xi_A)=(1/\sqrt3,-1)$ for
$\mhalf=900,600,300,250\gev$. Very little of the Yukawa-allowed
parameter space survives for the lowest choice. In the left-hand
window we fix $\mhalf=400\gev$ and vary $(\xi_0,\xi_A)$.
For the no-scale choice of $(0,0)$ very little of parameter space
yields an acceptable $\staupone$.

\FIG\mzeromin{We plot the minimum value of $\xi_0$ that is allowed
by the constraint $\mstaupone>{\rm max}\left\{45\gev,\mcnone\right\}$
for a given value of $\mhalf$ after scanning over all possible values
of $A$, assuming running masses $\mt=165\gev$,
$\mtaup=\mnup=50\gev$, $\mbp=\mtp=100\gev$,
and $\tanb=1.5$.}
Of course, the $\mstaupone>{\rm max}\left\{45\gev,\mcnone\right\}$
constraint is significant even if we
adopt a large (but allowed) value of $\mbpr$ and
do not employ a specific scenario for $\xi_0$ and $\xi_A$.  This
is illustrated in Fig.~\mzeromin. There, we take ($m(m)$ masses)
$\mt=165\gev$, $\mtp=\mbp=100\gev$, $\mtaup=\mnup=50\gev$
and $\tanb=1.5$ (for these mass
choices, solutions are only allowed for $\tanb$ values
very near 1.5) and plot
the smallest possible value for $\xi_0=m_0/\mhalf$ that is allowed as a
function
of $\mhalf$ after scanning over $\xi_A$ in the range $[-3,+3]$.
A very important generic feature emerges from the steep rise
of the minimum $\xi_0$ value as $\mhalf$ decreases:
for given $\mtp,\mbp,\mtaup,\mnup$ there is
a definite lower bound on $\mhalf$ (for reasonable values
of $\xi_0$) arising from the $\mstaupone>
{\rm max}\left\{45\gev,\mcnone\right\}$ constraint.  For the $\tpr$ and $\bpr$
masses considered, this bound is $\mhalf\gsim 140\gev$.
If we recall that
$\mgl\sim (\alphaz/\alpha_U)\mhalf$, and that $\alphaz/\alpha_U\sim 1.4$,
we see that this $\mgl$ bound translates into
a significant lower bound of $\mgl\gsim 200\gev$
deriving purely from limits on the fourth-generation $\staupone$ mass.
This type of constraint does not arise in the $\ng=3$ MSSM.

The ultimate lower bound on $\mhalf$ (along with the corresponding
lower bound on $\mgl$)
is actually quite independent of $\mbpr$. For instance,
if $\mbp$ is lowered to $50\gev$, keeping $\mt$ and $\mtp$
fixed, at $\xi_0=5$
the lowest allowed value of $\mhalf$ decreases by only about $5\gev$
compared to the $\mbp=100\gev$ value of $\sim 140\gev$ illustrated
in Fig.~\mzeromin. This is because,
at the ultimate lower bound, $m_0$ is very large and the $m_{H_1}^2$
term in ${\cal D}_{\taup}$ (see above) is swamped by the $m_{\staup_R}^2$
and $m_{\staup_L}^2$ terms which start off of order $m_0^2$ and remain large.

We re-emphasize the fact that in a specific SUGRA scenario with a fixed
value of $\xi_0$ (and of $\xi_A$) the lower bound on $\mhalf$ (and $\mgl$) can
be much larger than the high-$\xi_0$ ultimate lower bound if $\xi_0$
is small, as is illustrated in Fig.~\mzeromin.
(Of course, whatever the lower bound
on $\mgl(\mgl)$, $\mgl(pole)$ will be roughly 5-6\% higher.)
We also note that at fixed $\xi_0$
there is some dependence of the lower bound on $\mhalf$
upon the value of $\xi_A$.  This dependence is extremely weak at high $\xi_0$,
where $\xi_A$ values in the entire $[-3,+3]$ range generally give
an allowed solution for the lowest acceptable $\mhalf$ value.  But for lower
$\xi_0$ values, the lower bound on $\mhalf$ is generally achieved
only for $\xi_A$ values near 0. (In fact, the appropriate $\xi_A$ range
more or less scales with the magnitude of $\xi_0$.)

\FIG\glbhl{We plot the minimum possible value of $\mgl(\mgl)$,
and the corresponding values of $\mhl$ and $\mbp(pole)-\mhl-\mb$,
as a function of $\xi_A$ for a series of $\xi_0$ values:
$\xi_0=0$ (solid); $\xi_0=0.5$ (long dashes); $\xi_0=1$ (dots);
$\xi_0=2$ (dot-dash); $\xi_0=3$ (short dashes); $\xi_0=5$ (dash-dot-dot).
We have taken $\mt(\mt)=165\gev$,
$\mbp(\mbp)=\mtp(\mtp)=100\gev$ (corresponding to pole masses
of approximately $175\gev$ and $105\gev$). The minimum $\mgl$ is that
allowed by the $\staupone$ constraint for
a given $\xi_0,\xi_A$ choice. The corresponding $\tanb$ value
is either 1.5 or 1.6 in all cases. At each
$\xi_0$ and $\xi_A$ value, all values of $\mhalf\leq 10\tev$
were scanned. Curves terminate when no consistent solution is found.}
Perhaps one other plot
is useful in fully understanding the $\staupone$
constraint. In Fig.~\glbhl\ we consider various quantities
as a function of $\xi_A$ (ranging from $-3$ to $+3$)
for fixed  values of $\xi_0$ in the range from 0 to 5. In all cases
we take $\mbp(\mbp)=\mtp(\mtp)=100\gev$ (corresponding to pole
masses of approximately $105\gev$). For
each $\xi_0,\xi_A$ choice, we determine the minimum value of $\mhalf$
(the scan being confined to the region $\mhalf\leq 10\tev$)
that is consistent with the $\staupone$ constraint.
We plot the (minimum) value of $\mgl(\mgl)$ corresponding
to this minimum value of $\mhalf$.
(For later phenomenological use, we also
plot the corresponding $\mhl$ and $\mbp(pole)-\mhl-\mb$ values.)
For curves of limited extent in $\xi_A$,
the termination point(s) define the range beyond which
consistent solutions are not found with $\mhalf\leq 10\tev$.
The inconsistencies that arise at large $\xi_A$ are
of two types: (i) that the EWSB
solution requires $\mha^2<0$; and/or (ii)
that a 4th-family sparticle with color or charge
must have $m^2<0$, thereby breaking the color and/or U(1)-electromagnetic
symmetries.

This plot defines precisely the minimum $\mgl$ value that one can
have for a given $\xi_0,\xi_A$ boundary condition choice.  Once again
we see an absolute lower limit of order $\mgl(\mgl)\gsim 200\gev$
for our choice of $\mbp,\mtp$; the value of $\tanb$ for which
the minimum $\mgl$ value is reached is always in the range 1.5 to 1.6.
Note that for low $\xi_0$, the allowed solution range for $\xi_A$
is limited, and that the minimum $\mgl$ achievable increases substantially
when $\xi_A$ is not near 0.

\bigskip
\leftline{\bf 6. The Sparticle Spectrum}

We shall find in Sec.~7 that the first evidence for a fourth family
is very likely to be discovery of the $\bp$, $\tp$, $\taup$ and/or $\nup$
at LEP-II and/or the Tevatron. If one or more of these
fourth-family members are found,
the immediate question will be how this impacts the supersymmetric
particle spectrum, especially in the standard renormalization group
equation (RGE) context.
It is this latter issue that we address in this section.
First, we highlight the main features of the sparticle spectrum
in a four-family scenario, and compare the results to those obtained in
the three-family case. As previewed in Sec.~4,
we find that it is more than likely that some of the fourth generation
squarks and/or sleptons will be lighter than their counterparts
in the first three generations, with the $\staupone$ most probably
being the lightest. However, testing consistency
of their masses with unification and the RGE's is likely to be challenging
given the possibly large $A$-term induced mixing, and the probability
that they will have strange decay patterns and be tricky to observe.
Thus, we will focus primarily on gaugino masses and on the masses
of the sleptons and squarks of the first two generations,
focusing on when and how correlations among these masses will
be indirectly sensitive
to the presence of a fourth generation through the RGE's.
Certain relationships between masses are rather insensitive to
whether $\ng=4$ or $\ng=3$, and thus provide a test of the general RGE
context and universality of boundary conditions, while other mass
correlations are very different depending upon the value of $\ng$.
These latter relationships with strong $\ng$ dependence would
provide indirect evidence for the presence of a fourth generation, even if no
particle or sparticle belonging to the fourth generation is directly observed.

We first present some sample mass spectra, then discuss
mass sum rules and relations,
and finally focus on a specific correlation between
the first-family slepton masses and the LSP mass
that could reveal the presence of a fourth generation.

\bigskip

\leftline{\it 6a. Sample Mass Spectra}

For our illustrations we shall adopt $\mu>0$,
$\mtp=\mbp=100\gev$, and $\mtaup=\mnup=50\gev$.
We scan over the allowed
$\tanb$ values at any given choice for $\mgl(\mgl)$ and plot mass
spectra in units of $\mgl(\mgl)$. Three $\mgut$ boundary conditions
will be considered:
\pointbegin the dilaton scenario, with
$\xi_A\equiv A/\mhalf=-1$, $\xi_0\equiv m_0/\mhalf=1/\sqrt 3$;
\point the no-scale scenario, with $\xi_A=\xi_0=0$; and
\point the high-$m_0$ scenario, with $\xi_A=-1$, $\xi_0=1$.

\noindent
We begin by focusing on the spectrum for the $\staupone$ for these three
models.

\FIG\stauponefig{Mass spectra for the $\staupone$ for the three
standard $(\xi_0,\xi_A)$ choices. At the lower values of $\mgl$
the spectrum terminates either because $\mstaupone$ falls below
$45\gev$ (high-$m_0$ scenario), or because $\mstaupone<\mcnone$
(dilaton and no-scale scenarios). We have scanned in $\tanb$,
fixing $\mt(\mt)=165\gev$, $\mnup=\mtaup=50\gev$ and $\mtp=\mbp=100\gev$.}
In Fig.~\stauponefig, we plot the $\mstaupone$ spectra obtained by scanning
over allowed $\tanb$ values at various $\mgl$ values.  The lower
limit on $\mstaupone$ is fixed by $\mstaupone>45\gev$ in the case
of the high-$m_0$ scenario, and by $\mstaupone>\mcnone$ in the no-scale
and dilaton models.  The dramatic decrease of $\mstaupone$ as $\mgl$
decreases, discussed in Sec.~5, is evident. Of course, at high $\mgl$
values, the value of $\mstaupone$ is given roughly by $\mstaupone^2\sim
m_0^2+\hat C_{\staupone}\mhalf^2\propto (\xi_0^2+\hat C_{\staupone})\mgl^2$,
which is larger for larger values of $\xi_0$.

Let us now turn to the typical mass spectra for all the other supersymmetric
particles.  We begin with results for dilaton boundary conditions:
\FIG\spectrumdilaton{Mass spectra for dilaton boundary conditions. Masses
are given in units of $\mgl$.
We have taken $\mtaup=\mnup=50\gev$,
$\mtp=\mbp=100\gev$, $\mu>0$, and scanned in $\tanb$.}
Figure~\spectrumdilaton\ shows our results.
These can be compared to the $\ng=3$
results from Ref.~[\bgkp] for this same boundary
condition choice.\foot{In making comparisons, it is necessary to
note that for $\ng=3$ the value of
$\alpha_3=0.12$ was employed in Ref.~[\bgkp],
as compared to $\alpha_3\sim 0.128$
or so for our full two-loop four-generation treatment here.
Since $M_1:M_2:M_3\sim \alpha_1:\alpha_2:\alpha_3$
(all at $\mz$), $\mcnone$, $\mcntwo$, $\mcpone$ masses at a given $\mgl$
in Ref.~[\bgkp] for $\ng=3$
are approximately $0.94\sim 0.12/0.128$ times those appearing in
Fig.~\spectrumdilaton.}

The first big difference between the $\ng=4$ and $\ng=3$ cases
is the much larger lower bound on $\mgl$ in the former case.  As discussed
previously, this reflects the
$\mstaupone>{\rm max}\left\{45\gev,\mcnone\right\}$
requirement that is violated for low $\mgl$ values. A less dramatic
difference is that the allowed range of $\tanb$ is smaller for $\ng=4$,
and thus the value of $\mgl$ almost completely fixes the chargino
and neutralino masses, whereas for $\ng=3$ there is some scatter.
The limitation of $\tanb$ to low values for $\ng=4$ also means that (unlike
for $\ng=3$) the $\stauone$
remains almost degenerate with the $\slepr$'s of the first two families;
thus, a separate plot for the $\stauone$ is not given in our figures.
At a given $\mgl$, the most significant $\ng=4$ {\it vs.} $\ng=3$ difference
is the much larger masses for the squarks and sleptons
of the first three generations. Finally, there is the simple fact that
fourth-generation squarks and sleptons are present for $\ng=4$.
We have already noted
that $\staupone$ and $\snupone$  tend to be the lightest of the sleptons.
{}From Fig.~\spectrumdilaton\ we see that the lightest squark is very likely
to be the $\sbpone$. This is a rather general result.

As already hinted and more directly demonstrated in the next subsection,
the large squark and slepton masses for the first three
generation members can be directly
traced to the much larger value of $\alpha(\mgut)$ for $\ng=4$.
The basic idea is that
$m_{i}^2=m_0^2+\hat C_i\mhalf^2+D_i\mz^2\cos2\beta$,
where $i$ indicates the squark or slepton in question and
$\mhalf=\mgl\alpha(\mgut)/\alpha_3(\mgl)$.  Since
the $\hat C_i$ are not terribly different
in the three- and four-generation cases, the much larger $\alpha(\mgut)$
in the $\ng=4$ case greatly increases the squark or slepton mass
{\it at given $\mgl$}.
Combining this effect with the higher lower bound
for $\mgl$ in the $\ng=4$ {\it vs.}
$\ng=3$ case results in a large increase in the lower bounds on
squark and slepton masses in going to $\ng=4$.
For example, $\mslepl,\msnu,\mslepr\gsim 250,240,200\gev$
compared to $\mslepl,\msnu,\mslepr\gsim 110,52,45\gev$
in the $\ng=4$ and $\ng=3$ cases, respectively.
The high lower bound values in the $\ng=4$ case place these
sleptons from the first two families beyond
the reach of LEP-II and almost beyond the reach of a $\sqrt s=500\gev$
NLC. We shall see that the large squark and slepton masses
relative to gaugino masses
might well provide the most compelling indirect indication
for the presence of four generations
that one can obtain using only particles and sparticles
belonging to the first three generations.

\FIG\spectrumnoscale{Mass spectra for no-scale boundary conditions.}
In Fig.~\spectrumnoscale\ we give the corresponding results assuming no-scale
boundary conditions: $\xi_A=\xi_0=0$.  The first noteworthy
point is the much higher lower bound on $\mgl$ that arises in the
no-scale case, as compared to dilaton boundary conditions, when $\ng=4$.
Due to the zero value
of $m_0$ at $\mgut$, the slepton masses are smaller than
for dilaton boundary conditions, and, in particular, $\mstaupone$
is more easily driven to too low a value.
In addition, for the lower $\mgl$ values allowed by the $\mstaupone>\mcnone$
constraint, the $\snupone$ can be the LSP; it is often
very substantially lighter than the $\cnone$.
However, the $\slepl$ and $\snu$ masses
for the first two families are still much bigger than
found in the comparison results for $\ng=3$, see Ref.~[\bgkp].
The $\slepr$ is significantly lighter than the $\slepl$ in the no-scale
scenario (whatever $\ng$) because the associated soft-mass-squared
evolution is fed only by the $U(1)$ gaugino mass
terms, and moves to positive values (starting at $\mgut$ from 0 for $\xi_0=0$)
much more slowly than the soft-mass-squared that contributes
to the $\slepl$ and $\snu$ masses, which is fed by $SU(2)$ as
well as $U(1)$ gaugino mass terms (see Sec.~6b).
Squark masses remain very similar
to the dilaton scenario results due to the fact that
the $\hat C_{\wtil q}$ terms dominate the $m_0$ terms for
strongly-interacting sparticles. As in the dilaton case,
the large squark and slepton masses relative to $\mgl$ provide
a signal for $\ng=4$. Again there is a large
difference between $\ng=4$ and $\ng=3$ in the lower bounds so crucial for LEP
and NLC phenomenology. For example,
$\mslepl,\msnu,\mslepr\gsim 265,260,150\gev$
compared to $\mslepl,\msnu,\mslepr\gsim 110,60,70\gev$ for $\ng=4$ and $\ng=3$,
respectively. Thus, for $\ng=4$ only the $\slepr$ would be within
the kinematical reach of a $\sqrt s=500\gev$ NLC.

\FIG\spectrumhighm{Mass spectra for the large $m_0$ boundary condition case.}
In Fig.~\spectrumhighm\ we give results assuming the high-$m_0$ boundary
condition case: $\xi_0=-\xi_A=1$.  Here, the $\staupone$
constraint is more easily satisfied and $\mgl$ can take on lower values.
Even at the lowest allowed $\mgl$ values, the $\snupone$ is heavier
than the $\staupone$ (see Fig.~\stauponefig), which in turn is required to
be heavier than the LSP;
in fact, even at low $\mgl$ the $\snupone$ (but not the $\staupone$)
is heavier than the $\cpone$.
Squarks and sleptons receive a significant fraction of their mass from
the large $m_0$ value, and are thus more similar in mass.
However, the $\hat C_{\wtil q}$ contributions are still very important,
and squark and slepton masses for members of the first two families
continue to be much larger for $\ng=4$ than for $\ng=3$.
To repeat our previous sample comparisons,
$\mslepl,\msnu,\mslepr\gsim 245,240,220\gev$ for $\ng=4$
compared to $\mslepl,\msnu,\mslepr\gsim 120,89,107\gev$ for $\ng=3$.

\bigskip

\leftline{\it 6b. Sum Rule and Sparticle Spectrum Tests
for $\ng=3$ {\it vs.} $\ng=4$}

In the models considered here, having universal soft-SUSY-breaking scalar
mass $m_0$, there are many sum rules relating the
gaugino, squark and slepton masses.
Here, we survey the impact of a fourth family on the masses
and mass sum rules for
the gauginos and for the squarks and sleptons belonging to the
first two families;
\Ref\martram{S. Martin and P. Ramond, \prdj{48}, 5365 (1993),
provide a convenient summary of squark and slepton mass relations
in the $\ng=3$ case.}\
we also point out several differences between the sum
rules that relate only fourth-family masses and those that involve
only the third-family masses. Generally speaking, because
of the large mixings that can be present in both third and fourth
generation squark and slepton mass matrices, the most precise tests of
the consistency of RGE evolution and grand unification with sparticle
mass spectra may be those employing first and second generation members,
for which the mass matrices are very nearly diagonal.

To proceed, we must first discuss gaugino masses in more detail.
We employ the notation $k=1,2,3$ for the $U(1)$, $SU(2)$, and $SU(3)$
groups; $g_k(t)$ are the associated running coupling constants,
and $M_k(t)$ are the associated running gaugino masses,
where $t={1\over 2\pi}\log[Q(\gev)]$ --- $\tgut$ denotes $t$ at $Q=\mgut$,
$\tz$ denotes $t$ at $Q=\mz$, and so forth.
If the gaugino masses do indeed take a universal value $\mhalf$ at
$\mgut$ then at one-loop we have the standard relations:
$$M_k(t)=[\alpha_k(t)/\alpha(\tgut)]\mhalf\,;\eqn\mkgut$$
see Eqs.~(A.28) through
(A.31) in Appendix A. For an approximate idea of the numerics
for $\ng=4$ as compared to $\ng=3$, we take $\alpha_{em}^{-1}(\mz)=128$,
$\sinsq=.2316$ and $\alphaz=0.132$  at $t=\tz$ and $\alpha(\tgut)=0.0917$
(0.043) for $\ng=4$ ($\ng=3$).  This gives:
$$\pmatrix{M_1(\tz)\cr M_2(\tz)\cr M_3(\tz)\cr}
{\buildrel \ng=4\over =}\pmatrix{0.185\cr 0.37\cr
1.44\cr} \mhalf\,;
\qquad \pmatrix{M_1(\tz)\cr M_2(\tz)\cr M_3(\tz)\cr}
{\buildrel\ng=3\over =}\pmatrix{0.39\cr 0.78 \cr 3\cr}\mhalf\,.
\eqn\gauginomasses$$
These results are slightly modified when two-loop Yukawa contributions
to the running of the gauge coupling couplings are incorporated.
One finds coefficients for $\ng=4$ of roughly 0.194, 0.386 and 1.47
for $k=1$, 2 and 3, respectively.
The main effect derives from the simple fact that
$\alpha(\tgut)$ decreases so that the $\alpha_i(\tz)/\alpha(\tgut)$
ratios increase. For example, for $\mtp=\mbp=100\gev$,
$\alpha(\tgut)$ is shifted down by about 5\% to about 0.0874.
This results in an increase of the $\ng=4$, $k=1,2$
coefficients in Eq.~\gauginomasses\ by about 5\%;
for $k=3$, $\alpha_3(\tz)$ also decreases (to about 0.1288)
and the net Eq.~\gauginomasses\ coefficient increase is only about 2\%.

We also recall that simple asymptotic results for
the low-energy masses of the gauginos $\wtil\chi_i^0$ and $\wtil \chi_i^+$
arise if $\mw\ll ||\mu|\pm M_2|$ (but not if $\mw\sim ||\mu|\pm M_2|$).
These results are summarized in the chargino mass formulas
and neutralino mass formulas given in Eqs.~(36) and (37)
(for charginos) and (40), (41), (42) and (43) (for neutralinos) of
\REF\ghino{J.F. Gunion and H.E. Haber, \prdj{37}, 2515 (1988).}
Ref.~[\ghino]. (See also Eqs.~(5.3) and (6.8) of Ref.~[\martram].)
An example of such asymptotic results are the large $\mu$ relations
$\mcnone\sim M_1$ and $\mcpone\sim\mcntwo\sim M_2$,
where $M_{1,2}$ are evaluated at energy scales of order $\mz$.
As seen from Figs.~\spectrumdilaton, \spectrumnoscale\ and \spectrumhighm,
the above approximate mass formulae work reasonably well, but
not perfectly, at large $\mgl$. For instance, at high $\mgl$
one finds $\mcnone/\mgl\sim 0.135$ compared to $M_1/M_3\sim 0.131$
(with $\mtp=\mbp=100\gev$ two-loop effects included in the gauge running).

One of the first priorities of a next linear $\epem$ collider (NLC)
and the LHC will be to test the $\mgut$-scale
universality assumption for the $M_i$'s.  At the NLC it is estimated that
\REF\dpfsusy{See H. Baer \etal,
``Low Energy Supersymmetry Phenomenology'',
preprint FSU-HEP-950401, and references therein.}
$\mcpone$, $\mcnone$ and $\mcntwo$ can be measured to within a few
$\gev$.\refmark\dpfsusy\  This will allow a pretty good determination
of $M_1(\tgut)/M_2(\tgut)$.  The optimal situation
arises if $\mu$ is large, as
is essentially always the case in three-generation models, and very often
the case in the four-generation models. Then, as summarized
above (and making the approximation that
$\mcnone$ and $\mcpone$ are of order $\mz$)
$\mcnone\sim M_1(\tz)$ and $\mcpone\sim\mcntwo\sim M_2(\tz)$ and we have
the more general result:
$${M_1(\tgut)\over M_2(\tgut)}={\alpha_2(\tz)\over\alpha_1(\tz)}
{M_1(\tz)\over M_2(\tz)}\sim {\alpha_2(\tz)\over\alpha_1(\tz)}
{\mcnone\over \mcpone}\,.\eqn\universalitytest$$
(For greater accuracy, one will wish to include the two-loop corrections
to this relation when actually performing this test.)
Our ability to test
the universality for $i=3$ will probably be much more limited.
At the LHC, $\mgl$ is unlikely to be measured to better than 50 to 100 $\gev$.
Further, due to the small beta function at one-loop for $i=3$
in the case of four generations, two-loop corrections to the running
of $\alpha_3$ can be significant, and these depend at least somewhat
upon other model parameters.  Nonetheless, it should be possible
to extract a $\pm (20-30)\%$ value for $M_2(\mgut)/M_3(\mgut)$.
The determination of the $M_i(\tgut)$ is crucial, not only as a test
of universality, but also in making predictions for the squark and
slepton masses and testing $m_0$ universality.

But before turning to squarks and sleptons we must make a few
more comments on the gaugino sector. First, the universality test
will not have any significant sensitivity to $\ng=4$ {\it vs.}
$\ng=3$. Second, although the absolute
mass scales are sensitive to the limits on $\mhalf$ discussed earlier
that keep $\mhalf$ from being as small for $\ng=4$ as it can be
for $\ng=3$, it is difficult to use the absolute
mass scale for a reliable probe of the value of $\ng$ since
$\ng=3$ models with $\ng=4$ type $\mgl$ values are certainly entirely viable.
Sum rules such as
$$
\mcpone\mcptwo=|M_2\mu-\mw^2\sin2\beta|\,,
\eqn\charginorelation$$
following from the determinant of the mass matrix, are
sensitive to the $\ng=4$ restriction $1\lsim \tanb\lsim 3$ if
$M_2\mu\sim\mw^2$. But such small values are not the norm,
and to use this particular sum rule we must detect both $\cpone$ and $\cptwo$.
Thus, we must include squark and slepton masses in our considerations
in order to gain sensitivity to $\ng$.

We shall see below that
the squark and slepton masses at one-loop are determined by
the $m_0$ soft scalar mass (or masses should the $m_0$ values
not be universal), by $\beta$ and by the functions:
$$C_k(t)\mhalf^2\equiv \pmatrix{{3\over 5}\cr {3\over 4}\cr {4\over
3}\cr}\times {1\over \pi}\int_{t}^{\tgut}\,dt\,g_k^2(t)M_k^2(t)\,.
\eqn\ckforms$$
If one-loop evolution is used, then the integrals can be analytically
carried out to give:
$$C_k(t)\mhalf^2=2\pmatrix{{3\over 5}\cr {3\over 4}\cr {4\over 3}\cr}
M_k^2(t)\left[1-{\alpha^2(\tgut)\over\alpha_k^2(t)}\right] b_k^{-1}\,
\quad{\rm with}\quad b_k=\pmatrix{9-2\ng\cr 5-2\ng\cr -{3\over 5}-2\ng\cr}
\eqn\ckti$$
where the $b_k$ are the one-loop beta function coefficients for the gauge
couplings: $d\alpha_k/dt=-b_k\alpha_k^2$.
If we input the relations of Eq.~\mkgut, then (at one-loop) we find:
$$C_k(t)=2\pmatrix{{3\over 5}\cr {3\over 4}\cr {4\over 3}\cr}
\left[{\alpha_k^2(t)\over \alpha^2(\tgut)}-1\right] b_k^{-1}\,.\eqn\cktii$$
However, for four generations, one-loop evolution for $\alpha_3$ is generally
not an adequate approximation when calculating $C_3$; the
evolution of $\alpha_3$ is sensitive to two-loop terms,
due to the small value of $b_3$, and
$\alpha_3$ appears to the third power in the expression
for $C_3$ in Eq.~\ckforms.
The values of $C_k$ for $\ng=4$, computed numerically with full
two-loop evolution for the couplings, are compared to those for $\ng=3$
\TABLE\cktable{}
in Table~\cktable\ at scales $\mz$ and $1\tev$.
For the $\ng=4$ computations we adopted $\mtp=\mbp=90\gev$,
$\mtaup=\mnup=50\gev$, and $\tanb=1.5$.  Results are fairly
insensitive to these choices. In the case of $\ng=4$,
the results for $C_k$ would have been roughly 20\% bigger
had we neglected two-loop Yukawa terms in $\alpha_3$, which
keep $\alpha_3$ somewhat smaller (as described earlier) than
otherwise.

\midinsert
 \noindent{\tenpoint
 Table \cktable: We tabulate the $C_k$ values for $\ng=4$ and $\ng=3$
at the $\mz$ and $1\tev$ energy scales.
For $\ng=4$, we employ $\mtp=\mbp=90\gev$,
$\mtaup=\mnup=50\gev$, and $\tanb=1.5$.}
 \smallskip
\def\tstrut{\vrule height 12pt depth 4pt width 0pt}
 \thicksize=0pt
 \hrule \vskip .04in \hrule
 \begintable
$k$ |  $C_k(\tz)$; $\ng=4$ | $C_k(t_{1\tev})$; $\ng=4$
    |  $C_k(\tz)$; $\ng=3$ | $C_k(t_{1\tev})$; $\ng=3$ \cr
 1 | 0.131 | 0.130 | 0.151 | 0.148 \nr
 2 | 0.376 | 0.371 | 0.484 | 0.459 \nr
 3 | 4.59 | 4.06 | 7.30 | 5.40 \endtable
 \hrule \vskip .04in \hrule
\endinsert

We see a useful feature of the $\ng=4$ results in Table~\cktable:
the $C_k(t)$ are much more independent of the low-energy scale choice
than in the $\ng=3$ case.
This can be understood from Eq.~\cktii.  If we make a change in $\alpha_k(t)$,
then one can easily compute that:
$${\Delta C_k(t)\over C_k(t)}
\sim {2\Delta\alpha_k(t)\alpha_k(t)\alpha^{-2}(\tgut)\over
\left[{\alpha_k^2(t)\over \alpha^2(\tgut)}-1\right]}\,.\eqn\ckchange$$
First, consider $k=1,2$. Referring back to Fig.~\unification, we see that
to a reasonable approximation $\Delta \alpha_{1,2}$ for $\ng=3$
and $\ng=4$ are very similar in magnitude in going from $\mz$ to $1\tev$.
Also from Fig.~\unification, we see that for $k=1,2$ the denominator
of Eq.~\ckchange\ is very substantially
smaller, while $\alpha_k(\mz)/\alpha(\tgut)$ is much larger,
for $\ng=3$ compared to $\ng=4$. In combination, these two effects
lead to negligible change in $C_{1,2}$ in going from $\mz$ to $1\tev$
when $\ng=4$, compared to a modest change for $\ng=3$.
For $k=3$, the denominator is somewhat
larger for $\ng=3$ than $\ng=4$, but the numerator is very much smaller
for $\ng=4$, due not only to the decrease of $\alpha(\tgut)^{-2}$
by a factor of $\sim 4$, but also because $\Delta\alpha_3$
(in going from $\mz$ to $1\tev$) is very much smaller
for $\ng=4$ than for $\ng=3$ (see Fig.~\unification).
A substantial change in $C_3$
for $\ng=3$ is turned into a small change in the $\ng=4$ case.
This is amusing both in its own right, but is particularly important
for squark masses in that the small scale-sensitivity of $C_3$ means
the exact low-energy scale down to which we evolve will have a weak
impact upon the squark masses.
In the case of $\ng=3$,
when determining the running mass of a heavy squark, it is
crucial to evaluate $C_3$ at
the squark mass scale; for a heavy squark $C_3(t_{\wtil q})$ can
be very substantially smaller than the $\mz$-scale value as seen
in Table~\cktable. In contrast, even for $\ng=3$, $C_{1,2}$ are
much less sensitive to a change in scale.

\TABLE\fikditable{}
\midinsert
 \noindent{\tenpoint
 Table \fikditable: We tabulate the $f_{ik}$ and $D_i$ values for squarks and
sleptons. Also given are the $\mz$-scale $\hat C_i$ values for both
$\ng=3$ and $\ng=4$. For $\ng=4$, we employ $\mtp=\mbp=100\gev$,
$\mtaup=\mnup=50\gev$, and $\tanb=1.5$. We use the notation $\xw\equiv\sinsq$.}
 \smallskip
\def\tstrut{\vrule height 12pt depth 4pt width 0pt}
 \thicksize=0pt
 \hrule \vskip .04in \hrule
 \begintable
\ |  \multispan{3} \tstrut\hfil $f_{ik}$ \hfil | $D_i$ | $\hat C_i(\ng=4)$
| $\hat C_i(\ng=3)$ \cr
 $i$ \\ $k$ | 1       & 2 & 3 | \
| \    | \    \cr
 $\snu$   | ${1\over4}$ & 1 & 0 | ${1\over 2}$
| 0.41 | 0.52 \nr
 $\slepr$ | 1         & 0 & 0 | $-\xw$
| 0.13 | 0.15 \nr
 $\slepl$ | ${1\over4}$ & 1 & 0 | $-{1\over 2}+\xw$
| 0.41 | 0.52 \nr
 $\sdnr$  | ${1\over9}$ & 0 & 1 | $-{1\over 3}\xw$
| 4.43 | 7.32 \nr
 $\supr$  | ${4\over9}$ & 0 & 1 | ${2\over 3}\xw$
| 4.48 | 7.37 \nr
 $\sdnl$  | ${1\over36}$ & 1 & 1 | $-{1\over 2} +{1\over 3}\xw$
| 4.80 | 7.79 \nr
 $\supl$  | ${1\over36}$ & 1 & 1 | ${1\over 2} -{2\over 3}\xw$
| 4.80 | 7.79
 \endtable
\vskip .01in
 \hrule \vskip .04in \hrule
\endinsert

Let us now return to the squark and slepton masses.
The primary sensitivity of these masses to $\ng$ is
through the $C_k$ and the $\ng=4$ constraints on $m_0$ and $\mhalf$.
To see this let us recall that, in terms of the $C_k$, one finds
$$m_i^2=m_0^2+\hat C_i\mhalf^2+D_i\mz^2\cos2\beta\,,
{}~~{\rm where}\quad \hat C_i \equiv \sum_k f_{ik}C_k\,.\eqn\miforms$$
The $f_{ik}$ and $D_i$ are tabulated in Table~\fikditable, as are
$\tz$-scale values for the $\hat C_i$, for both $\ng=3$ and $\ng=4$;
in evaluating the $\hat C_i$'s we have employed
full two-loop evolution for
all of the $\alpha_k$'s in determining the $C_k$'s, as
noted earlier. Also noted above is the fact that, for $\ng=4$,
employing the $\mz$-scale values for the $\hat C_i$
should be a reasonably good approximation,
even in the case of heavy squarks, due to the modest sensitivity
of the $C_k$ coefficients to the energy scale.
Of course, we must keep in mind that the numbers given above have
presumed the universal value $M_i=\mhalf$ at $\mgut$.  Clearly,
if the $M_i$ do not have a common $\mgut$-scale value, one would
have to redo all the computations.

What about the value of $\beta=\tan^{-1} v_2/v_1$?
Whether or not the slepton and squark masses are measurably
sensitive to $\beta$
depends upon the relative size of the $m_0^2$ and $\hat C_i\mhalf^2$
terms compared to $\mz^2\cos2\beta$. Certainly it would be nice
to have sensitivity, since then it could be determined
if $\tanb$ falls in the $1\lsim \tanb\lsim 3$ domain required if $\ng=4$.
There are two mass-squared differences that, in principle,
allow a direct determination of $\cos2\beta$ independent of whether
the $m_0$'s are universal and independent of any knowledge
of the numerical values of the $\hat C_i$:
$$\mslepl^2-\msnu^2=\msdnl^2-\msupl^2=-\cos2\beta \mz^2(1-\xw)\,,\eqn\deli$$
where we employ the notation $\xw\equiv\sinsq$.
(Note that since $\tanb\geq 1$, $\cos2\beta$
is always negative and these mass differences are positive.)
For such differences to be sensitive to $\cos2\beta$ at just the
$1\sigma$ level, one must be able to measure a typical $\wtil m$
to accuracy $\Delta \mtil /\mtil <\mw^2/2\mtil^2$.
For masses of $100,250,500,1000\gev$,
this means $\Delta \mtil/\mtil < 32\%,5\%,1.3\% ,0.32\%$;
clearly the last two
accuracies are at a {\it very} difficult level
even for an $\epem$ or $\mu^+\mu^-$ collider of adequate energy.
Determination of $\cos2\beta$ at the $3\sigma$ level of accuracy
would require factor of three smaller errors than those listed above.
To be more precise regarding the possibilities
for measuring the mass differences in Eq.~\deli,
we outline some relevant issues and possible techniques
in Appendix B.

The conclusion from Appendix B is that it will be difficult
to measure $\mslepl$ and $\msnu$ to much better than about 5\% accuracy.
(Since the mass scales for
$\msdnl$ and $\msupl$ are even larger than for the sleptons, and errors
in the determination of their masses will be larger also (see Appendix B),
it seems clear that it is the slepton sector upon which we should focus.)
Given the 5\% accuracy estimate for $\mslepl$ and $\msnu$ determination
and comparing to the criteria of the previous paragraph,
we conclude that a direct determination
of $\cos2\beta$ will be problematical,
especially if the slepton mass scales are as large as predicted for $\ng=4$.
The ability to determine $\tanb$ could be even worse if $\tanb$ is large,
since $\cos2\beta$ varies slowly with $\beta$ once $\beta$ approaches
$\pi/2$.  Of course, for $\ng=4$ it is true that
$\tanb$ is in the $1\lsim \tanb\lsim3$ range
where our ability to extract $\tanb$ from a $\cos2\beta$ measurement
would be maximal.  Nonetheless, for the moment we must conclude that
experimental determination of $\cos2\beta$ and thence $\tanb$
will be difficult unless the slepton masses are well below $200\gev$,
as is possible only if $\ng=3$.  Of course,
further study is undoubtedly warranted and could reverse this conclusion.
Fortunately, the uncertainty
in $\cos2\beta$ is not the limiting factor in our ability
to test the other crucial mass sum rules and relations discussed below.
Other experimental and theoretical uncertainties
are much more important in determining the limitations.

\medskip
\undertext{Sum rules that do not mix slepton/squark and gaugino masses}
\medskip

We can imagine two basic outcomes
for our experimental attempts to determine $\cos2\beta$: i) we are successful
in getting at least a rough determination of $\mz^2\cos2\beta$; or ii)
our errors are too large to be sensitive to this kind of term.
In either case, one
would proceed next to a series of additional sum rules and mass relations
that would provide tests of universality and the general unification/RGE
framework and/or tests for $\ng=4$ {\it vs.} $\ng=3$.
For the moment, let us assume that both squarks
and sleptons are observable, if not at an $\epem$ collider then
perhaps at a $\mu^+\mu^-$ collider of very high energy.\Ref\bbgh{V. Barger,
M. Berger, J. Gunion and T. Han, preprint UCD-95-12 (1995).}
If not, one must proceed quite differently; we will return to this alternate
case in Sec.~6c.

To check $m_0$ universality independently of $\ng$,
one must consider mass combinations that are independent of the $C_i$.
The only simple example is
$$\eqalign{&2(\msupr^2-\msdnr^2)+(\msdnr^2-\msdnl^2)+(\mslepl^2-\mslepr^2)\cr
&={10\over3}\xw\mz^2\cos2\beta=
{10\over 3}\xw(1-\xw)^{-1}(\msnu^2-\mslepl^2)\,.\cr}\eqn\univtest$$
This relation would be violated if all the $m_0$'s appearing
in Eq.~\miforms\ are not the same. Because of the
large number of mass differences appearing above,
to detect a violation of universality in just one of these differences
at the $\Delta m_0/m_0=f$ level could require that the squark and slepton
masses be measured with an accuracy that is $\lsim f/4$.
{}From the discussion of Appendix B, we conclude that
it will be difficult to obtain accuracy $f$ that is much better than $15-20\%$.
However, this level of accuracy is acceptable
in the sense that we will find that tests of all other sum rules are more or
less restricted to this same rough level of accuracy even
if the $m_0$ were universal. This means that the limited accuracy of
the test of $m_0$ universality
will not dominate our ability to check various mass relations.
Let us presume that universality is shown to be satisfied within
the accuracy of measurement achievable.
Then we can proceed to extract $m_0$ and test for $\ng=3$ {\it vs.} $\ng=4$
to about this same level of accuracy.

There are various mass combinations that can be used to extract a
universal $m_0$. However, we should
note that the four quantities $m_0$, $C_1$, $C_2$, and $\mz^2\cos2\beta$
cannot be extracted using just the three slepton masses ---
we must have some squark-sector measurements. One simple combination is
$$m_0^2=\mslepr^2-3(\msupr^2-\msdnr^2)+4\xw(1-\xw)^{-1}(\msnu^2-\mslepl^2)\,.
\eqn\mextract$$
Note that the $C_3$ terms that cancel between $\msupr^2$ and $\msdnr^2$
will be evaluated at very similar mass scales, so that the cancellation
should be quite precise. The $C_1$ terms that cancel between
$\mslepr^2$ and $-3(\msupr^2-\msdnr^2)$ could in principle be
evaluated at somewhat different mass scales.  As discussed
previously, this would not matter in the $\ng=4$ case, but would
lead to a small uncertainty for $\ng=3$; however, this uncertainty is
much less than that from simple experimental errors
in determining $\msupr^2-\msdnr^2$.  Indeed,
if $m_0$ is small, then experimental
errors in evaluating the right-hand side above could become a
severe problem. We will not dwell further on this issue here.

The primary sensitivity to $\ng=4$ {\it vs.} $\ng=3$ derives
from the fact that the gauge coupling at unification, $\alpha(\tgut)$,
is approximately twice as large when $\ng=4$ as it is when $\ng=3$.
This leads to the possibly large difference between the $\ng=4$
and $\ng=3$ values of $C_3$,
depending upon the energy scale of evaluation, illustrated in Table~\cktable.
However, it is not straightforward to exploit this sensitivity
through mass  relations involving squarks and sleptons alone,
that is relations designed to eliminate direct reference to $\mhalf$.
This is already apparent from the modest changes in the $C_k$ and $\hat C_i$
(see Tables~\cktable\ and \fikditable) in going from 3 to 4 generations.
The very best that one can do is to consider ratios such as $ C_3/C_1$.
As seen from Table~\cktable, if squarks are at a low mass
scale (of order $\mz$) then this ratio takes the values 33.5 for $\ng=4$
as compared to 48.3 for $\ng=3$; but if squarks have masses of order $1\tev$,
then we compute $ C_3/ C_1=36.5$, not very different from
the 33.5 value for $\ng=4$.
Experimental extraction of $C_1$ and $C_3$ can be done in a number of ways.
An example is:
$$\eqalign{
C_1\mhalf^2=&3(\msupr^2-\msdnr^2)-3\xw(1-\xw)^{-1}(\msnu^2-\mslepl^2)\,\cr
C_3\mhalf^2=&(\msupl^2-\msnu^2)+{2\over 3}(\msupr^2-\msdnr^2)
\,,\cr}\eqn\ciciiiforms$$
where the $C_1\mhalf^2$ result was already used in Eq.~\mextract.
Accuracy of about 10\% or better would be very desirable
for the experimental determination of $C_1\mhalf^2$ and $C_3\mhalf^2$
if we are to have a good chance of distinguishing
$\ng=4$ from $\ng=3$, but
would require determination of the $\msupl-\mslepl$ and $\msupr-\msdnr$
mass differences to roughly 3\%, a challenging
task even at an $\epem$ collider, as we have already discussed.

Some mass sum rules are quite insensitive to $\ng=3$ {\it vs.} $\ng=4$.
Such mass relations can be used to test the general RGE approach,
somewhat independently of the generation issue.
To illustrate, consider the sample relation (3.19) in Ref.~[\martram]:\foot{
We have corrected a sign error in this equation.}
$$\eqalign{
m_{\wtil e_L}^2 -m_{\wtil e_R}^2=&
{(C_2-{3\over 4}C_1)\over (C_3-{8\over 9}C_1)}
\left(m_{\wtil d_R}^2-m_{\wtil e_R}^2\right) \cr
+&\cos 2\beta\mz^2\left[-{2\over 3}\xw
{(C_2-{3\over 4}C_1)\over (C_3-{8\over 9} C_1)}-({1\over
2}-2\xw)\right].\cr}
\eqn\sumrulei$$
We note that $\cos2\beta<0$ (for $\tanb>1$)
so that the sign of the second term is opposite to the
sign of the factor in brackets. Adopting our results from Table~\cktable\
for the $C_k$, we find at scale $\mz$:
$$
m_{\wtil e_L}^2-m_{\wtil e_R}^2=(0.052~{\rm or}~0.062) (m_{\wtil d_R}^2-
m_{\wtil e_R}^2)- [(19.3 \gev)^2~{\rm or}~(19.6 \gev)^2]\cos2\beta\,,
\eqn\mrestimate$$
for $\ng=3$ or $\ng=4$, respectively.  Since this
relation assumes universality among the $m_0$, the relatively minor changes
in the coefficients for $\ng=4$ compared to $\ng=3$ implies greater sensitivity
to possible universality violation among the $m_0$ than to $\ng$.
We note that for accurate examination of this sum rule, an
accurate value for $C_3$ is absolutely necessary.  As noted above,
in computing $C_3$ it is
critical when $\ng=4$ to include the two-loop contributions to the running
of $\alpha_3$ given that the one-loop beta function is anomalously small,
being proportional to $9-2\ng$.

Third-generation mass sum rules (4.5) and (4.10) of Ref.~[\martram],
which are independent of the $m_0$ universality assumption,
are unaffected by a fourth family.  However, since they require that the
$\wtil b$ mass matrix is diagonal to a good approximation (which is
generally true for the third family since $\mb$ is small), the analogous
sum rules do not apply for the fourth family due to the large
value of $\mbp$.

A generalization of the sum rule (4.6) of Ref.~[\martram]
for the splitting $m_{\wtil
t_1}^2-m_{\wtil t_2}^2$ can be found in the four-family case.  It is
not especially useful, however, since it relates
$m_{\wtil \tpr_1}^2-m_{\wtil \tpr_2}^2$
to $m_{\wtil \bpr_1}^2-m_{\wtil \bpr_2}^2$ via $A_{\tpr}$ and $A_{\bpr}$
dependent terms; thus, we do not display it here.

\bigskip

\undertext{Sum rules that mix squark/slepton masses and gaugino masses}
\medskip

So far, we discussed only mass relations and sum rules that
referred either to squark/slepton masses only, or to gaugino masses only.
However, it should be clear from all our previous discussions
that the real sensitivity to $\ng=4$ {\it vs.} $\ng=3$ lies in
the dramatic shift of slepton/squark masses at given $\mgl$.
This was already discussed in Sec.~6a, and is further evident
from the relations of Eq.~\ciciiiforms, once the substitution
$\mhalf=\mgl\alpha(\tgut)/\alpha_3(\mgl)$ is made and it is recalled
that $\alpha(\tgut)|_{\ng=4}=0.092$ compared to $\alpha(\tgut)|_{\ng=3}=0.043$.
However, because of inherent inaccuracies in experimentally
measuring $\mgl$ it is better to reference $M_1(\tz)$ or $M_2(\tz)$,
see Eq.~\gauginomasses,
which can be quite well determined once the experimentally
more accessible masses $\mcnone$ and $\mcpone$ are measured.
A particularly simple example is provided by $\msdnr^2-\mslepr^2$.
{}From Table~\fikditable\ we find that (using $\mz$-scale $\hat C_i$ values ---
for $1\tev$ scale values the difference in coefficients is even
greater) and the relations of Eq.~\gauginomasses,
$$\eqalign{\msdnr^2-\mslepr^2
=&\delta m_0^2+(4.3~{\rm or}~7.17)\mhalf^2+{2\over 3}\xw\mz^2\cos 2\beta\cr
=&\delta m_0^2+(31.5~{\rm or}~11.8)M_2^2(\tz)+
{2\over 3}\xw\mz^2\cos 2\beta\,,\cr}
\eqn\mixi$$
for $\ng=4$ or $\ng=3$, respectively. Here, we have included a
possible violation of $m_0$ universality, $\delta m_0^2\equiv
m_0^2(\sdnr)-m_0^2(\slepr)$. Experimental uncertainties
in determining the coefficient, $C_{M_2}$, of $M_2^2(\tz)$ will be dominated
by experimental errors in determining $\msdnr^2-\mslepr^2$ and
$\delta m_0^2$. The latter error, as discussed with regard to
Eq.~\univtest, is expected to be several times as large as the former.
Very crudely, for experimental uncertainties
in the squark/slepton mass measurements of size $\Delta \mtil/\mtil\sim f$,
one expects an experimental uncertainty
in the $M_2^2$ coefficient of order
$\Delta C_{M_2} \sim 3-5\times f \mtil^2/M_2^2$.
We have already seen that $f$ is very unlikely to be better than $f\sim
0.1$. Taking $\mtil\sim 1\tev$ and $M_2\sim 200\gev$ implies
a $\Delta C_{M_2}\sim 7.5-12.5$.  This is smaller than the predicted
$\ng=4$ {\it vs.} $\ng=3$ difference, suggesting we would, in fact,
be sensitive to $\ng$ by measuring masses of the gauginos and the squarks
and sleptons of just the first two generations.

\bigskip
\leftline{\it 6c. Non-Squark Spectrum Tests for Universality and
Four Generations}

As repeatedly noted, mass determinations will be most easily performed at
an $\epem$ collider of appropriate energy.
Regardless of the value of $\ng$, chargino pair production
is very likely to be within reach of a $\sqrt s=500\gev$ NLC.
In the case of $\ng=3$, slepton pair production is also very likely
to be possible for $\sqrt s=500\gev$ if boundary conditions
of the no-scale or dilaton type are appropriate.  For $\ng=4$, however,
these same boundary conditions predict much larger slepton masses,
and a higher energy machine would probably be required.  However,
for either $\ng=3$ or $\ng=4$, the squarks are much heavier still,
and it is quite possible that an $\epem$ or $\mu^+\mu^-$
collider would be built
that could produce the neutralinos, charginos and sleptons,
but not the gluino and squarks.
Thus, it is desirable to consider what kinds of tests of universality,
the RGE framework, and $\ng$ can be performed without reference to either
the gluino or the squarks.
In particular, we wish to determine if there is a test for $\ng=3$ {\it vs.}
$\ng=4$ that can be performed using only the
$\cnone,\cpone$ masses and the masses of the first or second
family sleptons, which are those most likely to be both easily accessible
and most precisely predictable (\ie\ without reference to
$A$ parameters and $L-R$ mixing). So, let us imagine
that the $M_i$ for $i=1,2$ have been determined with reasonable
accuracy from the observed values of $\mcnone$ and $\mcpone$,
following the procedure described earlier. If we then assume
a value of $\ng$, $\mhalf$
will also be computable from Eq.~\gauginomasses.
Finally, the $C_i$ can also be computed for the assumed value of $\ng$.

As mentioned earlier, the $\slepl$ and $\slepr$ masses are very sensitive
to $\ng$ when expressed in terms of the parameter $\xi_0$.
In what follows, we keep track of a possible difference between the $\xi_0$
values in the $L$ and $R$
sectors, \ie\ of possible universality violation.
We have, from Table~\fikditable,
$$\eqalign{
\mslepl^2\sim &[\xi_0^L]^2\mhalf^2+(0.41~{\rm
or}~0.52)\mhalf^2+(-\half+\xw)\mz^2\cos2\beta\,,\cr
\mslepr^2\sim &[\xi_0^R]^2\mhalf^2+(0.13~{\rm
or}~0.15)\mhalf^2-\xw\mz^2\cos2\beta\,,\cr}
\eqn\mlrform$$
for $\ng=4$ or $\ng=3$, respectively.
We note that the $\hat C_i$ for $\mslepl$ and $\mslepr$ are quite
different, but not terribly sensitive to $\ng=3$ {\it vs.}
$\ng=4$.  However, by converting to the
value of, for example, $M_1(\tz)$
using Eq.~\gauginomasses, a large difference between the $\ng=4$ and
$\ng=3$ phenomenology emerges.
In the approximation that $\mcnone\sim M_1(\tz)$ we have
$$\eqalign{
\rl^2\equiv {\mslepl^2\over \mcnone^2}\sim &(29~{\rm or}~7)[\xi_0^L]^2+(12~{\rm
or}~3)+(-\half+\xw){\mz^2\over\mcnone^2}\cos2\beta\,,\cr
\rr^2\equiv {\mslepr^2\over \mcnone^2}\sim &(29~{\rm or}~7)[\xi_0^R]^2+(4~{\rm
or}~1)-\xw{\mz^2\over\mcnone^2}\cos2\beta\,,\cr}
\eqn\mlrratios$$
for $\ng=4$ or $\ng=3$, respectively.

We plot our precise numerical results for
$\rl$ and $\rr$ as a function of $\xi_0$
\FIG\discriminate{A plot of $\rl$ and $\rr$ (see text)
as a function of $\xi_0$ for the typical case of $\mcnone=117\gev$.
We have taken $\tanb=1.5$, $\mt=165\gev$,
and, in the $\ng=4$ case, $\mtp=\mbp=100\gev$, $\mtaup=\mnup=50\gev$.}
in Fig.~\discriminate. We have taken $\mt=165\gev$
and $\tanb=1.5$. In the case of $\ng=4$, we choose $\mtp=\mbp=100\gev$
and $\mtaup=\mnup=50\gev$. (All masses are running masses in this discussion.)
The graph is given for the specific value of
$\mcnone=117\gev$. For $\ng=4$ (3) this corresponds to
$\mhalf\sim 600\gev$ ($\sim 300\gev$), respectively.
This kind of plot has a number of important advantages.  First,
there is no sensitivity to the $A$
parameter for the first and second generation masses being considered.
Second, $\tanb$ enters into the location of the $\xi_0=0$ intercepts,
but only weakly if $\mcnone\sim\mz$. Third,
by measuring the $\slepl$ and $\slepr$
masses in units of $\mcnone$, the plot is quite independent
of the actual value of $\mhalf$ (or, equivalently, $\mcnone$),
as seen in the approximate relations, Eq.~\mlrratios.

Fig.~\discriminate\ exhibits a rather big difference
for the functional dependence of $\rl$ and $\rr$ on $\xi_0$,
depending on $\ng$.  Suppose that experiment yields $\rr=4.5$.
For $\ng=3$ this implies $\xi_0^R\sim 1.7$ and, if $\xi_0^L=\xi_0^R$,
\ie\ the $\xi_0$ are universal,  $\rl\sim 4.7$;
for $\ng=4$ we find $\xi_0^R\sim 0.8$ and $\rl\sim 5.3$.
Since $\mcnone\sim 117\gev$ in this case, this corresponds
to $\mslepl(\ng=4)-\mslepl(\ng=3)\sim 70\gev$ compared to perhaps $10-20\gev$
uncertainty in the $\slepl$ mass measurement.  Of course, to measure
$\mslepl$ and $\mslepr$ in the mass range above $400\gev$,
required in this example, demands a very high energy $\epem$
or $\mu^+\mu^-$ collider.

As one moves to lower $\xi_0$ values, the discriminating power
of this procedure slowly increases.
Suppose the experimentally measured value is
$\rr=3$.  If $\ng=3$ this would imply $\xi_0^R=1.1$, and $\xi_0^L=\xi_0^R$
would then imply $\rl=3.4$.  In contrast,
if $\ng=4$, $\rr=3$ corresponds to $\xi_0^R=0.45$, yielding
$\rl=4.1$, or $\mslepl(\ng=4)-\mslepl(\ng=3)\sim 82\gev$, a bit
larger than our previous result.

How sensitive are we to a breaking of universality and the uncertainty
in $\tanb$? Consider first the possibility that $\xi_0^L$
and $\xi_0^R$ are not the same.  Returning
to our first example, and neglecting effects
of the $Tr(Ym^2)$ evolution terms, a measurement of $\rl=5.3$
in agreement with the $\ng=4$ prediction without universality violation
could be reinterpreted as a prediction of $\ng=3$ provided $\xi_0^L\sim 1.9$,
as compared to the value $\xi_0^R=1.7$ required for $\rr=4.5$ for $\ng=3$.
This translates to a breaking of universality in the amount:
$${\xi_0^L-\xi_0^R \over {1\over 2} (\xi_0^L+\xi_0^R)}
\sim {0.2\over 1.8}\sim 11\%\,.\eqn\ubreak$$
Thus, $\sim 10\%$ universality violation could correct for
the $\ng=4$ {\it vs.} $\ng=3$ difference.  This is an intrinsic
ambiguity when only slepton, chargino and neutralino
masses are available.

In the above, we have purposely chosen situations where consistent
solutions for both $\ng=3$ and $\ng=4$ are possible.
It could easily happen that consistent choices for $\xi_0^R$ and $\xi_0^L$
are simply not possible for the observed values of $\rr$ and/or $\rl$.
In particular, if $\rr\lsim 2.0$
or $\rl\lsim 3.5$ then $\ng=4$ is excluded, whereas $\ng=3$ models
would be possible. In fact, if $\ng=3$ and $\xi_0$
is in the preferred $\xi_0\lsim 1$ range, then $\rl<3$ and $\ng=4$
would be immediately excluded.
The lower bounds on $\rl$ and $\rr$ for $\ng=4$ are, of course,
equivalent to lower bounds on $\mslepl,\mslepr$ of $410,234\gev$
for this particular $\mcnone$ value.  And, these lower bounds can
only be reached if one is willing to accept the no-scale $m_0=0$
boundary conditions. In another extreme, it could happen that one measures
$\rr\sim\rl\gsim 7-8$.  Such large slepton masses relative to $\mcnone$
require very large $\xi_0$ values if $\ng=3$, well beyond any acceptable
model range. In contrast, if $\ng=4$ the required $\xi_0$ values
are perfectly reasonable.

We note that sensitivity of these results to
$\cos 2\beta$ is really quite limited. For $\mcnone\gsim\mz$,
and taking both $\xw$ and $-\half+\xw$ to be of magnitude $\sim 0.25$,
the full range of $\cos 2\beta$ between 0 and $-1$ corresponds to
$\Delta r\sim 0.125/r\times \mz/\mcnone\lsim 0.125/r$. Since the $r$
values are typically at least of order $1$, most probably much larger,
we see that the $\cos2\beta$ uncertainty is, at worst, roughly
the same as the experimental error, and is not significant compared
to the predicted $r$ differences or possible universality questions.

Finally, we note that even if we have an ambiguity in the
decision between non-universality and $\ng=3$ {\it vs.} $\ng=4$, in general
we will have a very rough determination of the general size of $\xi_0$.
In the above example, we would at least know that the $\xi_0$ values
were $\gsim 0.8$ if $\ng=4$ or $\gsim 1.7$ if $\ng=3$.  This
would allow us to set the overall boundary condition picture.

\vskip 1in
\bigskip
\leftline{\bf 7. Experiment and a Fourth Family}

In this section, unless otherwise indicated, all masses are {\it pole}
masses.  Where appropriate, running masses will be explicitly
written in the $m(m)$ notation. Very
roughly, pole masses are higher than running
$m(m)$ masses by about 5 or 6\%, based on the dominant correction
$m(pole)=m(m)[1+{4\over 3\pi}\alpha_s(m)]$.

In all our discussions up to this point, we have
adopted the conventional interpretation that the CDF
and D0 events derive from $t\rta b W$ decays, with $\mt(pole)\sim 175\gev$.
Alternative possibilities in which the events at CDF and D0 are
reinterpreted as $\mtp\gsim 175\gev$ events with $\tp\rta b W$,
while the $t$ is not currently observed,
are not consistent within the perturbative MSSM GUT framework.
In the first such option, the $t$ is not observed because it is too heavy,
$\mt(\mt)\gg 165\gev$. The difficulty with this option is that
for such large values of $\mt$
the maximum value of $\mtp(\mtp)$ allowed by perturbativity is
much below $165\gev$ and the $\tp\bar\tp$ events would not have
the same characteristics ({\it e.g.} reconstructed mass and cross
section) as observed by CDF and D0. In the second such option
(emphasized in Ref.~[\habertop]), the $t$ is much lighter than the $\tp$,
$\mt\sim\mw$, but is not detected because its decays are dominated
by $t\rta {\wt t_1}\cnone$. However, in the MSSM GUT context, the latter decay
does not occur if the soft-SUSY-breaking parameters are approximately
universal, since (as we have seen)
the $\wt t_1$ is always much heavier than the $t$.  Further discussion
of this latter case will appear in Sec. 7a. Certainly, the phenomenology of
the $\tp$ and $\bp$ at the Tevatron that we present depends crucially
on the $\mt\sim 175\gev>\mbp,\mtp$ hierarchy, not to mention
the relative size of $\mtp$ {\it vs.} $\mbp$,
and on the mixing pattern between the generations.

In Sec.~7a we will discuss direct limits on the fourth family fermions
coming from collider experiments, especially the CDF and D0 experiments
at the Tevatron, assuming that $\mt(pole)\sim175\gev$.
We find that a relatively small
portion of $\mbp,\mtp$ parameter space survives all such
constraints for the more natural choices of CKM mixing
between the fourth generation and lower generations.
We also describe in more detail why the $\mt\sim\mw$ scenario
is not viable for universal soft-SUSY-breaking boundary conditions.
In Sec.~7b we give the constraints on $\mtp,\mbp$ arising from
consistency with precision electroweak data.  Not surprisingly,
these tend to rule out large $\mtp$ compared to $\mbp$.
Finally, in Sec.~7c we speculate on deviations in single-jet inclusive
and di-jet spectra that could arise as a result of a full four-family
content at relatively low mass scales.

\bigskip
\leftline{\it 7a. Direct Collider Searches}

As we have demonstrated, the requirement that all Yukawa couplings remain
perturbative places rather strict upper bounds on the
$\tpr $, $\bpr$, $\nupr$ and $\taupr$ masses, and the possibility of an MSSM
fourth generation may well be settled within the next few years by searches
(i) at the Tevatron, for an appropriate set of new signatures (\ie\ besides
$t\rightarrow b+W$), and (ii) at LEP-II, via detection
of direct $\taupr +\bar \taupr $ and/or $\nupr +\bar \nupr $ production.

LEP has set the firmest direct limit, namely $m_f\gsim \mz/2$
for $f=\bpr,\tpr ,\taupr ,\nupr $. We have also seen that
the value of $\mt\sim 175\gev$, as determined by the recent
CDF and D0 results,\refmark\newtopresults\
is such as to imply $\mt>\mbp,\mtp$ for perturbative consistency.

Let us begin by noting the critical difference between the
$\mtp<\mbp$ and $\mtp>\mbp$ cases, assuming that $\mt>\mtp,\mbp$.
If $\mtp<\mbp$, then the $\tp\rta Wb$ decay proceeds at tree-level
at a rate determined by the CKM entry $V_{\tp b}$.
One-loop processes, such $\tp\rta c g$, $\tp\rta c \gamma$,
$\tp \rta c Z^{(*)}$ and so forth occur at rates determined by
$V_{\tp s}V_{cs}$ and $V_{\tp\bp}V_{c\bp}$ times a loop integration factor.
It would be extremely unnatural for the product of a loop integration
factor times a two-generation-skipping CKM matrix element to exceed
the one-generation-skipping CKM $V_{\tp b}$.  Thus, it is almost certain
that $\tp\rta Wb$ decays would be completely dominant if $\mtp <\mbp$.
This situation should be contrasted with the reverse case of $\mtp >\mbp$,
for which the interesting issue is how the $\bp$ decays.
The decay $\bp\rta W c$ occurs at tree-level with rate determined
by $V_{c \bp}$. The one-loop rates for $\bp\rta b g$, $\bp\rta b \gamma$,
$\bp \rta b Z^{(*)}$ and so forth are determined by
$V_{t\bp}V_{tb}$ and $V_{\tp \bp}V_{\tp b}$ times loop integration factors.
It is actually
rather likely that the two-generation-skipping CKM element $V_{c\bp}$
is smaller than the one-generation-skipping
$V_{t\bp}$ or $V_{\tp b}$ times the loop integration factor.  In this
case, the flavor changing decays $\bp\rta b X$ would
dominate over the two-generation skipping $\bp\rta W c$ decay.
The different decay structure for $\mtp<\mbp$ as compared to $\mtp>\mbp$
has crucial phenomenological implications.

To repeat, if $\mtp<\mbp$ it is almost certain that the $\tp$
will decay in a top-quark-like manner
to $W b$ (or $W^* b$, depending on $\mtp$).  Events
of this type are clearly ruled out
by early CDF and D0 `top-quark' limits for all $\mtp<\mbp$ regions
allowed by the four-generation perturbativity constraints, see
Fig.~\allowedregions. Thus, we immediately eliminate roughly
half of the perturbatively allowed regions. We now focus on
experimental restrictions that arise if $\mtp>\mbp$.

\medskip
\undertext{$\bp\bar \bp$ Production at the Tevatron}
\medskip

If $\mtp>\mbp$, then in $\bp\bar\bp$ production
at the Tevatron we must consider the two more or less
competitive $\bp$ decay scenarios outlined above. If
$V_{\bpr c}$ (or, much less likely, $V_{\bpr u}$) is sufficiently large
that $\bpr\rta c W$ (or $\bp\rta u W$) decays are dominant,
the published data
\Ref\abe{F. Abe \etal, \prlj{64}, 147 (1990), \prlj{68}, 447 (1992);
T. Trippe, Particle Data Group, private communication.}
already rule out $\mbp<85\gev$ at $95\%$ C.L. (Presumably
this old result will soon be updated using the full CDF and D0 data sets.)
But, if the $\bpr$ is essentially unmixed with light generations,
it will decay via flavor-changing neutral current (FCNC) channels:
$\bpr \rightarrow b \gamma,b g$ or
$bZ^*$ for $m_{\bpr }<\mz$, with $\bpr \rta bZ$
becoming dominant for $\mbp>\mz+\mb$.

Present data would appear to rule out
a significant rate for $\bp\rta b Z$ decays by virtue of there
not being room for an excess number of $Z$'s beyond that predicted
on the basis of the standard $q\anti q$ annihilation mechanism.
Taking a $\bp$ mass of $100\gev$ (roughly the maximum possible
if $\mtp>\mbp$), the cross section (without
inclusion of any $K$ factor) at $\sqrt s=1.8\tev$ (Fermilab
Tevatron) is about $60\pb$.  For an integrated luminosity of
$67\pbi$ this implies roughly 4000 $\bp\bar\bp$ events.
If we imagine looking for $Z$'s from the decay of either (or both) $\bp$'s,
then this number is to be multiplied by $[1-BR(\bp\rta {\rm non-}Z)^2]$.
Given the small phase space, $\mbp-\mz-\mb\sim 5\gev$
and the low value of $\mtp\sim 100\gev$ required
by our perturbative constraints (which impacts the one-loop
Penguin diagram calculations), we expect\refmark\AGRAWALetal\
$BR(\bp\rta {\rm non-}Z)$ to be significant but probably $\lsim 0.7$,
implying $\gsim 2000$ events in which at least one of the $\bp$'s decays
to a $Z$. Using $BR(Z\rta\ell^+\ell^-)=0.067$, and
an overall two-lepton cuts/acceptance/detection
efficiency of about $\epsilon=0.5$,
we find $\gsim 80$ events where at least one $Z$ appears in
the $\bp\bar \bp$ final state and has a clearly detectable leptonic decay.
(If we take $BR(\bp\rta Z b)=1$, and adjust for luminosity, this number
is consistent with the pre-$p_T(Z)$-cut numbers of Agrawal, Ellis and Hou,
Ref.~[\AGRAWALetal], for the sum of $\ell^+\ell^-+jj+X$ and $4\ell+X$ event
numbers at $\mbp=100\gev$.)
In addition, Fig.~\allowedregions\ shows that if $\mtp$ is to be
larger than $\mbp$ and $\mbp\sim 100\gev$, then $\mtp$ can only
be {\it just} larger than $\mbp$. In this case, the $\tp\bar\tp$
production cross section would be more or less the
same as the $\bp\bar \bp$ cross section, thereby almost doubling
the 80 events, assuming that $\tp\rta \bp W^*$ decay is dominant.
Although a specific analysis has not been
presented for the latest CDF or D0 data with regard to this point,
we expect that such a large number of additional relatively clean $Z$ events
can be ruled out. For an integrated luminosity of $L=150\pbi$, it would
seem that one could virtually eliminate the $\mbp>\mz+\mb$ case
down to very near the threshold. However, if $\bp\rta b\hl$ is also
kinematically allowed, the number of extra $Z$ events would
be much smaller and $\mbp>\mz+\mb$ would not be excluded.
We shall return to this issue shortly.

Thus, consistency with experiment probably requires that $\mbp<\mz+\mb$,
and that $\bp$ decays are dominated by the FCNC channels outlined above.
These `non-standard' decays would yield distinctly different
signatures than the lepton-plus-jet signatures of charged current decays
that are already mostly excluded, and new search strategies are required.
Perhaps the cleanest discovery channel makes use of $\bp\bar\bp$
production followed by double $\bp\rta b\gamma$ decay.
As an example consider $\mbp=90\gev(95\gev)$.  The Tevatron cross section
(without including any QCD $K$ factor enhancement)
for $\bp\bar\bp$ production is $\sim 100\pb(76\pb)$, corresponding
to $\sim 6700(5100)$ events for $L=67\pbi$. Assuming
$BR(\bp\rta b\gamma)\sim 0.1$ (typical of the low $\mtp$ results given
in Hou and Stewart, see Ref.~[\AGRAWALetal]), we obtain
about 67(51) $b\overline b \gamma\gamma$ events.
In order for these events to be sufficiently free of background,
single $b$ tagging is probably necessary.  If we assume that
the efficiency for tagging of
at least one $b$ (after appropriate kinematical cuts) is approximately
50\% (as found in the most recent CDF top-quark analysis\refmark\newtopresults)
and that the overall
cuts/acceptance/detection efficiency (after requiring minimum transverse
momentum, say $p_T>15\gev$, and central rapidity
for the photons and $b$'s)
is about 35\%, we see that a clean sample of 11(8) or so
events should be present in the CDF data. Of course, $\tp\bar\tp$ production
will also contribute to the $b\anti b \gamma\gamma X$
signal assuming $\tp\rta \bp W^*$ is the dominant decay of the $\tp$.
(For $\mtp$ near $\mbp$ the jets or leptons from the $W^*$ will be quite
soft, and the signal rather similar to that for direct $\bp\bar\bp$
production).
Events of the $2\gamma+2j$ type with at least one tagged $b$ quark
are currently under investigation by the CDF collaboration,
\Ref\HOUSTON{J. Houston and R. Blair, private communication.}\
and we expect that CDF will shortly be able to severely restrict
the allowed four-generation scenarios. Since event rates rise rapidly
with decreasing $\mbp$, it is hard to imagine that a $\bp$
with mass much below $80-90\gev$ would escape detection.
Referring to Fig.~\allowedregions,
this means that the $\mbp(\mbp)<75-85\gev$ region for $\tanb=1.5$ and the
entire $\tanb=2.2$ region are
on the verge of elimination if no signal is reported.
An integrated Tevatron luminosity of $L=150\pbi$ would surely
allow one to use the $b\anti b \gamma\gamma$ final state to eliminate
all $\mbp\lsim \mb+\mz$, extending up to $\mbp$ values such
that the $\bp\rta b Z$ mode really becomes significant enough
to suppress the $\bp\rta b\gamma$ branching ratio. At this point,
as described above, one looks for extra $Z$'s, which also provide an excellent
signal.

Of course, if a reasonable $K$ factor is included in the $\bp\bar \bp$
cross section ($K=1.3$ to 1.5), the event rates for the $b\anti b\gamma\gamma$
and/or extra $Z$ events will increase above those given above.
This leaves even less room for the four-family model to avoid detection.

The only possible escape from the $\bp\rta b \gamma$
and $\bp\rta b Z$ decay mode constraints would be
if $\bp\rta b \hl$ decay is kinematically allowed.  If allowed,
it will certainly dominate all other modes
when $\bp\rta b Z$ is kinematically forbidden. Even when the $\bp\rta b Z$
channel is open, $\bp\rta b\hl$ will at the very least be competitive
and suppress the $\bp\rta b Z$ branching ratio --- indeed, the $b\hl$ channel
could still be the dominant
mode.\Ref\BpHb{W.-S. Hou and R. Stuart, \prdj{43}, 3669 (1991).}
However, as seen in
Fig.~\higgsnewlimit\ (after converting running masses to pole masses),
this escape is not available for the
$\mtp>\mbp$ portion of parameter space in the case of the
dilaton boundary condition scenario, since there $\mbp<\mhl$.
Only in the extreme lower right-hand corner where the running masses obey
$\mhl\simeq \mbp\simeq 100\gev$
and $\mtp\simeq 50\gev$, is the $\bp\rta b\hl$ decay kinematically
allowed, and this corner is almost certainly ruled out (as discussed
in an earlier paragraph) since $\mtp<\mbp$.
However, for scenarios with larger (but not too large) $m_0$, the
$\hl$ {\it can} be light enough to make the $b\hl$ channel kinematically
accessible.

In Fig.~\glbhl\ we plotted as a function of $\xi_A$, for various
fixed $\xi_0$ values, the minimum $\mgl(\mgl)$ value that is
allowed by the $\staupone$ constraints, taking
$\mbp(\mbp)=\mtp(\mtp)=100\gev$ (corresponding to pole masses
of about $105\gev$).
We can now use this minimum
$\mgl(\mgl)$ value to fix all other model parameters,
thereby determining the sparticle spectrum, in particular $\mhl$.
Since the $m_{H_{1,2}}^2$ contributions to
the $m_i^2=m_{H_i}^2+\mu^2$ mass terms in the scalar potential
will tend to be smallest when $\mgl$ is small (and $\mu$
also tends to scale with $\mgl$),
these minimum $\mgl$ values give $\mhl$ values that
will be very close to the minimum allowed for the fourth-generation
fermion masses chosen. To determine if $\bp\to\hl b$
is kinematically allowed, we compute
$\mbp(pole)-\mhl-\mb$.  Plots of $\mhl$ and this mass difference
also appeared in Fig.~\glbhl. From the $\mbp(pole)-\mhl-\mb$
curves, we see that for middle-of-the-road values of $\xi_0$
the $\bp\rta b\hl$ decay can be kinematically allowed. When the
$\bp\rta b\hl$ decay channel is present, thereby suppressing
the $\bp\rta bZ$ channel branching ratio,  a $\bp$ with
pole mass somewhat above $105\gev$ could have
escaped observation.

\vskip .75in
\medskip
\undertext{$\tp\bar \tp$ Production at the Tevatron}
\medskip

Let us now turn to a more thorough examination of $\tp\bar\tp$ events.
For the $\mtp>100\gev$ mass region,
there are two competing decay modes for the $\tp$: $\tp\rta \bp W^{(*)}$
and $\tp\rta b W$. The $\bp W$ decay will certainly dominate
over the $b W$ mode when the former $W$ is real, since the $bW$ channel is
suppressed by the intra-generational mixing factor
$V_{\tp b}$.  However, when $\mtp-\mbp< \mw$, the two-body $bW$
mode could be competitive with the three-body $\bp W^*$ decay.
\FIG\tpdecayratio{We plot the maximum value of $V_{\tp b}$ for
which $\Gamma(\tp\rta \bp W^*)>\Gamma(\tp\rta b W)$ as a function
of $\mtp$ for fixed values of $\mbp=80\gev$ and $100\gev$.
All masses are pole masses.}
In Fig.~\tpdecayratio\ we plot the value of $V_{\tp b}$ for which
$\Gamma(\tp\rta \bp W^*)=\Gamma(\tp\rta b W)$ as a function of $\mtp$
for several $\mbp$ values (we employ pole masses).
For $V_{\tp b}$ larger than the values plotted,
the $\tp\rta b W$ mode is dominant. We see that for $\bp$ (pole)
masses larger than the rough lower bound of $80\gev$, discussed above,
it is not at all impossible that $V_{\tp b}$ could be
large enough for $\tp\rta b W$ to be the dominant mode.

For $\mtp\lsim 170\gev$ (as required by our Yukawa perturbativity bound),
such dominance would imply a large excess of events relative to those
already present by virtue of $t\anti t$ production and decay.
As discussed in more detail below, the CDF and D0 experimental
results are now more or less consistent with the expected $t\anti t$
rates for $\mt\sim 175-185\gev$. Taken at face value, these results then
imply that $\tp\rta b W$ cannot be the dominant $\tp$ decay mode.

In fact, we show below that it is only possible to obtain
sufficiently few CDF and D0 events if the $\tp$ decay is to $\bp$ plus
a highly virtual $W^*$. A typical case that survives is $\mtp\sim 115\gev$
and $\mbp\sim 80\gev$. From Fig.~\tpdecayratio\ we find that
$V_{\tp b}\lsim 0.02$ (a value roughly the same as $V_{cb}$)
is required for $\tp\rta b W$ to be adequately
suppressed for these mass choices. Such values for $V_{\tp b}$ are
certainly reasonable; in fact, one might expect considerably
smaller values based on the fact that $V_{bc}\ll V_{us}$,
but we must also consider implications for $\bp$ decay.
We noted earlier that if $\bp\rta c W$ were the dominant
$\bp$ decay mode, then $\bp\bar \bp$ events would have been detected.
Consistency with experiment thus
requires that the one-loop $\bp \rta b X$ neutral current decays are dominant.
This, in turn, requires (see Hou and Stewart, Ref.~[\AGRAWALetal])
that $V_{\tp b}/V_{\bp c},V_{t\bp}/V_{\bp c}\gsim 10^2-10^3$. Combining with
the $V_{\tp b}$ restriction above, we find that $V_{\bp c}\lsim 10^{-4}$
is required. As seen from Fig.~\tpdecayratio, this limit would become stronger
for larger values of $\mbp$.  Suppressions larger than those suggested
by measured $V_{cb}$ and $V_{ub}$ values would be required for
$\mbp$ values much above $100\gev$ and thus become problematical. As discussed
earlier, $\mbp\gsim 100\gev$
is also the range for which $\bp\rta \hl b$ decay
dominance is probably required to avoid too many extra $Z$ events
from $\bp\rta Z b$ decays in $\bp\bar \bp$ events.

In our previous work,\refmark\GMP\ we performed a rough Monte-Carlo study
of the number of $\tpr \bar \tpr $
production and decay events expected
in the dilepton-plus-jets and lepton-plus-jets
channels, assuming that $\tp\rta \bp W^{(*)}$
is the primary $\tp$ decay. At the time, the number of additional
events (beyond the predicted number from $t\anti t$ production)
in these two channels was not inconsistent with the then-available
CDF data for any of the larger $\mtp$, $\mbp$ choices within the perturbatively
allowed region.
However, the luminosity accumulated by both CDF and D0 is now
much larger ($L\sim 67\pbi$ {\it vs.} $19\pbi$).  Further, CDF now employs new
$b$-tagging algorithms which roughly double their $b$-tagging efficiency.
Consequently, the constraints on the model are now much stronger.
Exactly how severe the constraints are depends on the precise predictions
for event rates from normal $t\overline t$ production.  The CDF collaboration
states that their results are in good agreement with a top-quark Monte Carlo.
As a cross check, we have
also repeated the Monte Carlo study of our previous work, Ref.~[\GMP],
after adjusting for the new luminosity and new $b$-tagging
procedures. Despite the approximate nature of our implementation of the
CDF cuts and $b$-tagging procedures, agreement is good. For $\mt=175\gev$,
the uncut $t\anti t$ cross section is about $3.5\pb$, and roughly 25
events pass the single-lepton plus $b$-tag plus $\ge 3$-jets criteria,
while $\sim 3$ events pass the dilepton cuts.
CDF observes 37 events in in the $W+\geq 3$ jet channel in which at least
one $b$ is tagged, with an estimated background of 16, and 6 events
in the di-lepton channel, with estimated background of 1.3.
\foot{The number of background events, 16, in the $W+\ge 3$-jets channel
is estimated by scaling the tag background estimate of 22 events
by the ratio 37/50 of events/tags.}
Thus, their number of signal events is $21$ and $4.7$ in
these two respective channels, to be compared to 25 and 3.
Clearly, there is little room for additional events
passing their cuts from $\tp\bar \tp$ production.

\TABLE\tlikerates{}
\midinsert
 \noindent{\tenpoint
 Table \tlikerates: For $L=67\pbi$, we tabulate the number of $\tp\bar\tp$
events passing our approximations to the
CDF top-quark discovery cuts in the single lepton+$b$-tag and dilepton modes.}
 \smallskip
\def\tstrut{\vrule height 12pt depth 4pt width 0pt}
 \thicksize=0pt
 \hrule \vskip .04in \hrule
 \begintable
\ | \multispan{4} \tstrut\hfil Lepton + $b$-tag Mode \hfil |
\multispan{4} \tstrut\hfil Dilepton Mode \hfil \cr
\ | \multispan{4} \tstrut\hfil $m_{b'}$ (GeV)  \hfil |
\multispan{4} \tstrut\hfil $m_{b'}$ (GeV) \hfil \nr
$m_{t'}$ (GeV) | 50 & 80 & 110 & 130 | 50 & 80 & 110 & 130 \cr
  160 |  20 & 16 & 3.6 & 1.5 | 2.4 & 1.8 & 0.2 & 0.01 \nr
  130 |  53 & 20 & 0.5 & $-$ | 4.2 & 1.1 & 0.002 & $-$ \nr
  100 |  58 & 3.6 & $-$ & $-$ | 2.8 & 0.01 & $-$ & $-$
\endtable
 \hrule \vskip .04in \hrule
\endinsert

The rates (after implementing the same cuts, {\it etc.})
in these two channels deriving from $\tp\bar \tp$ production,
followed by FCNC decays of the type $\bp\rta b +jet(s)$
(in particular, no extra photons or leptons from $\bp$ decay are allowed)
appear in Table~\tlikerates. (In Table~\tlikerates\ we have included several
($\mtp,\mbp$) pairs that are excluded by the boundaries of
Fig.~\allowedregions\
just to indicate the effects of different mass choices on the rates.)
Restrictions on the perturbatively allowed regions of Fig.~\allowedregions\
are now significant.  For $\tanb=2.2$, for which only values of
$\mbp<85\gev$ ({\it i.e.} $\mbp(\mbp)\lsim 80\gev$ are perturbatively allowed,
Table~\tlikerates\ shows $\tp\bar\tp$ production
will yield sufficiently few CDF single-lepton + $b$-tag
events only if $\mtp\lsim 105-115\gev$, effectively eliminating
the upper 50\% of the allowed $\mtp(\mtp)$ region. For $\mtp\lsim 105-115\gev$,
Table~\tlikerates\ also shows that $\mbp$ can't be too light; roughly
$\mbp\gsim 70\gev$ is required.\foot{
Note that the $\tp\rta \wtil b^\prime_1 \cpone$ decay,
that could provide an escape from these restrictions, is not kinematically
allowed for any of the mass spectra scenarios of
Figs.~\spectrumdilaton-\spectrumhighm.}
The underlying reason for these restrictions
is easily summarized.  For $\mtp\lsim \mt$, the only way the $\tp\bar\tp$
events can evade being included in the CDF event sample is if $\mtp-\mbp$
is sufficiently small that the $W$ in $\tp\rta W \bp$
is virtual and the jets and leptons from the two $W^*$'s are quite soft.
For such cases, additional $\tp\bar\tp$ events
could be included in the event sample
if CDF and D0 could soften their cuts.  However,
backgrounds probably increase dramatically.
If this is the case, then additional
luminosity is the only way to improve sensitivity; for example, $L=150\pbi$
would yield about 8 CDF events (for current cuts and $b$-tagging efficiency)
if $\mtp=100$ and $\mbp=80$.  This would probably constitute a detectable
signal.

\medskip
\undertext{Combined Restrictions}
\medskip

Combining the $\tp$ restrictions with the existing and probable $\bp$
CDF data constraints, and accounting for
the limited range allowed for $\mbp$ by perturbativity (especially
for $\tanb=2.2$) and the $\mtp>\mbp$ requirement,
we are clearly forced into the {\it running} mass domain
$110\gev\gsim\mtp(\mtp)\gsim\mbp(\mbp)$, $95\gev\gsim \mbp(\mbp)\gsim 75\gev$.
For $\tanb=1.5$ this is the region that
is as close to a fixed-point limit as the perturbativity constraints
allow and is also the region for which the unification prediction for
$\alpha_s(\mz)$ is smallest and, thus, in best agreement with data.
We have also seen that it is a region for which $V_{\tp b}$
and $V_{\bp c}$ can both take on values that are reasonable extraplolations
from those already measured in the lighter quark sectors,
while at the same time being consistent with dominance of
$\tp\rta \bp W^*$ over (probably excluded) $\tp\rta b W$ decays
in $\tp\bar \tp$ events
and of $\bp\rta X b$ neutral current decays over (probably excluded)
$\bp\rta c W$ decays in $\bp\bar \bp$ events, {\it provided} $\mtp$
($\mbp$) lies in the upper (lower) portion of its allowed region.
A tripling of the Tevatron's integrated luminosity to $L\sim 150\pbi$
would almost certainly either close the $\mbp$ window above, or
lead to $\bp$ discovery.

\medskip
\undertext{The Hidden-Top Scenario}
\medskip

To escape from the  experimental constraints on $\tp\bar \tp$
production and decay discussed above
would require that the `top-quark' events
not come from third-generation $t\anti t$ production.
As described earlier, in one such scenario
$\mt\sim \mw$ and $t\rta \wt t_1 \cnone$ decays are dominant.
Typical (pole) masses required for this scenario to be viable are those
discussed in Ref.~[\habertop]: $\mt\sim \mw$,
$\mtp\sim 170\gev$, $\mbp\sim 110\gev$, $\mcnone\lsim 25\gev$,
and $m_{\wt t}\lsim 60\gev$.
We have explicitly analyzed this scenario using our full SUSY GUT framework.
We find that for $\mtaup=50\gev$ (the choice of Ref.~[\habertop])
and $\mnup\lsim 78\gev$, the above $\mt$, $\mtp$ and $\mbp$
masses are consistent with gauge coupling unification and with
perturbative evolution for the Yukawas up to the scale $\mgut$.
(Our criteria do not quite allow
consistency with Yukawa perturbativity for the $\mnup=80\gev$
choice of Ref.~[\habertop].)

However, we find that it is not possible to
obtain $\mcnone+m_{\wt t_1}<\mt$ (where $\wt t_1$ is the lighter
stop mass eigenstate) in the context of SUSY GUT scenarios
with universal $m_0$ and $A$ parameters at $\mgut$.
Since both $\mcnone$ and $m_{\wt t_1}$ scale with $\mgl$,
the minimum value for the mass sum will occur at the smallest
possible value of $\mgl$.  As described earlier,
the minimum $\mgl$ (or equivalently $\mhalf$)
value is set by the $\staupone$ constraint.
We find that the $\staupone$ constraint
(for the mass scenario just outlined) implies
an absolute lower bound on the $\mgut$-scale parameter
$\mhalf$ that is similar in nature to that illustrated
in Fig.~\mzeromin, except that lower values
of $\mhalf$  (close to $100\gev$) are allowed at high $\xi_0$.
This translates into $\mgl\gsim 150\gev$.
Analogously to Fig.~\glbhl,
much higher values are required when $\xi_0\equiv m_0/\mhalf\lsim 1$.
Also as in Fig.~\glbhl, the minimum value of $\mhalf$ is achieved for
a range of $\xi_A\equiv A/\mhalf$ values centered
about 0 which is broad for large $\xi_0$, but increasingly narrow
for $\xi_0\lsim 1$. However, even at large $\xi_0$, arbitrariy large
values of $|\xi_A|$ do not yield consistent solutions for
any $\mhalf$, as illustrated by the termination of
the low-$\xi_0$ curves in Fig.~\glbhl. Too large
a value for $|\xi_A|$ (typically $|\xi_A|/\xi_0\gsim 4-5$ at large $\xi_0$)
results in either $\mha^2<0$ at the EWSB minimum,
and/or $m^2<0$ for one of the colored and/or charged 4th-family sparticles.

We describe two typical cases: $\xi_0\sim 1$ and $\xi_0\sim 5$.
For $\xi_0\sim 1$, $\mgl\gsim 400\gev$ is required, with
$\mcnone\sim 0.125-0.13 \mgl$ implying $\mcnone\gsim 50\gev$.
For $\xi_0\gsim 5$, we find $\mgl\gsim 150\gev$ and $\mcnone\sim 0.14\mgl$,
implying $\mcnone\gsim 21\gev$. ( Both of these minimum values
only apply for small $\xi_A/\xi_0$. As $\xi_A/\xi_0$ increases,
the minimum allowed value of $\mgl$ increases, but very slowly
in the case of $\xi_0=5$.)
Although the latter $\mcnone$ value is in an acceptable range for the scenario,
$m_{\wt t_1}$ is always quite large for $\xi_A$ values that do
not imply $\mha^2<0$ and/or color/U(1) breaking. This is because
$m_{\wt t_1}$ is always well above $\mt=80\gev$ at small $\xi_A$
(typically by three hundred GeV or more for $\xi_0=5$). Thus, even though
$\wt t_L - \wt t_R$ mixing increases so that $m_{\wt t_1}$ decreases
as one moves to higher $|\xi_A|$ in the $\xi_A<0$ direction, $m_{\wt t_1}$
only declines by some 50 GeV before $|\xi_A|$
is large enough that $\mha^2<0$ and/or color/U(1) breaking arises.

For allowed $\xi_A$ values, what prevents a sufficiently small
mass for the $\wt t_1$
is the fact that the $\staupone$ cannot be too light. The simplest way
to avoid this lock-step arrangement is to break
the universality assumption for the $m_0$ parameters at $\mgut$,
allowing much smaller $m_0$ in the $\wt t$ sector than in the $\staup$
sector of the theory. We have not explored this option.

\vskip .5in

\leftline{{\it 7b. Four-Family Precision EW tests}
\foot{In this sub-section we will exclusively employ running masses.}}
\medskip

In addition to direct collider searches for the presence of a possible fourth
family, indirect effects arising from virtual quantum effects can be
probed through precision measurements of many observables and can lead to
indirect limits or bounds on the parameters of the theory. For example, the
well-known $\rho$-parameter limits the magnitude of iso-spin breaking, and
thus bounds the $\mtp^2-\mbp^2$ and $\mtaup^2-\mnup^2$ splittings of a possible
fourth family. On the other hand, although the LEP measurement
of $Z\rta b\bar b$
is a particularly sensitive probe to any new physics involving possible
coupling to the top quark, the vertex contribution to $Z\rta b\bar b$
arising from a fourth family is expected to be small due to CKM mixing
suppression. We therefore neglect all vertex contributions in our
analysis, focusing instead on vacuum polarization effects.

Overall, it desirable
to perform a global fit to the present experimental data using several
different observables. The most comprehensive approach would be to perform the
complete, one-loop calculation for each observable in question, and do a
best fit of the new physical parameters of the model, ({\it i.e.}
$\mtp$, $\mbp$, $\mtaup$, $\mnup$, $\tanb$, $\mu$, $\ldots$).
This rather ambitious approach
has only recently been attempted for the MSSM, and will not be attempted
here. A more modest approach inevitably involves making certain
assumptions and approximations. i) As previously explained, a
fourth-family added to the MSSM contributes dominantly via 'oblique',
or vacuum polarization effects. ii) We assume that the
contributions from the sparticles are small enough to be neglected.
This is likely to be a good approximation for $\ng=4$ models.
Indeed, we have seen that for $\ng=4$ the supergravity boundary
conditions and evolution equations imply a fairly large
lower bound on $\mgl$ (especially for the dilaton and no-scale
boundary condition choices) which, in turn, implies that most
of the sparticle masses are large.
According to the decoupling theorem, loops involving massive SUSY
particles contribute negligibly to vacuum
polarization amplitudes. The particles most likely to violate
this approximation are the sometimes-light $\staupone$ and $\snupone$,
and even these are heavy for larger $\mgl$ values, independent of model.
Finally, iii) we
assume that the $\hl$ of the Higgs sector is SM-like, implying
that the SUSY Higgs sector is equivalent to a SM Higgs sector with
a light Higgs boson. Given the four-family
constraint $\tanb\lsim 3$, this is true for $\mha\gsim 200\gev$, as
is essentially always the case for the models considered.

In the following analysis, we perform a global fit to the latest LEP+SLD
data\Ref\data{D. Schaile, talk presented at the {\it 27th Int. Conf. on
High Energy Physics,}, Glasgow, Scotland, July 20-27, 1994; SLD Collaboration,
F. Abe \etal, \prlj{73}, 25 (1994).} and we
employ an extension of the Peskin/Takeuchi S,T,U formalism
\Ref\peskin{M. Peskin and T. Takeuchi, \prdj{46}, 381 (1992).}
in order to find the $90\%,95\%$
allowed regions of parameter space. In this approach, for a given `reference'
SM ({\it i.e.} with chosen $\mt,m_H$ values), $S,T,U\equiv 0$, and the
(non-zero) best-fit values to experimental data
are a measure of the oblique, or vacuum polarization,
contributions that should be explained by
`new physics'. For our reference SM we adopt $\ng=3$ with
$\mt(\mt)=165\gev$ and Higgs mass $m_H=100\gev$ (typical of
the $\mhl$ values found in our computations). We then treat the
contributions from $\tpr,\bpr,\nupr,\taupr$ as new physics.

At this point, there are several parameter
choices and assumptions that affect our results. i) What
values of $\alpha_{em}(\mz),\alpha_s(\mz)$ are chosen
for the reference SM? As was recently
shown, new estimates of $\alpha_{em}(\mz)$ can shift the best-fit
for the experimental
value of the $S$-parameter relative to $S=0$ for the reference SM
by $+10\%$. Similarly, the choice of
$\alpha_s(\mz)$ is quite important. Secondly, ii) one should be careful when
using the $S,T,U$ formalism in the presence of
new light physics, since in this case it has recently been shown that
the standard $S,T,U$ formalism is simply {\it not} an
\REF\burgess{C. P. Burgess, S. Godfrey, H. Konig,
D. London and I. Maksymyk, \prdj{49}, 6115 (1994); C. P. Burgess, S.
Godfrey, H. Konig, D. London and I. Maksymyk, \plbj{326}, 276 (1994);
I. Maksymyk, C. P. Burgess and D. London, \prdj{50}, 529 (1994);
P. Bamert, C.P.Burgess MCGILL-94-27, (1994).}
\REF\roy{A. Kundu and P. Roy,
SINP-TNP/94-07, TIFR/TH/94-38.}
adequate parameterization.\refmark{\burgess,\roy}\
In the original work of Peskin and Takeuchi, the
$S,T,U$ parameters were explicitly defined using the linear approximation
in an expansion in $q^2/M^2_Z$.\refmark{\peskin}\
Although perfectly valid when considering
new physics with scales much higher than $\mz$, such as technicolor,
this expansion is not accurate for new physics near or below $\mz$. It
{\it is} possible to define a set of parameters which do
not rely on a $q^2/M^2_Z$ expansion.
In a complete comparison to all available experimental
observables, this results in a
proliferation of new parameters that must be simultaneously fit.\foot{In
a quadratic expansion, in general {\it seven} independent parameters arise
\refmark{\burgess,\roy},
and in a fully exact treatment, {\it eleven} parameters
must be specified\Ref\py{H. Pois, T.C. Yuan, unpublished}}
However,
if one restricts the `global' fit to Z-pole plus $m^2_W/m^2_Z$ measurements,
then one can show that a re-definition of $S\rta S',T\rta T',U\rta U'$ is
possible, where the primed variables can be computed in the presence
of additional light physics without approximation. Further all
the observables in question are linear functions of $S',T',U'$.
Thus, the procedure is to restrict the fit to measurables which
are defined at the Z-peak (along with the $m_W/m_Z$ measurement), employ
the standard three-parameter $S,T,U$ global fit to the
experimental data (relative to a reference SM with $S=T=U=0$
by definition), and finally re-interpret the best fit $S,T,U$ values
as $S',T',U'$.

In order to compare to predictions of a four-generation model,
we calculate the exact (no expansion is performed)
contributions to $S',T',U'$ from loops involving fourth-generation
particles for a given set of $\mt,\mtp,\mbp,\mnup,\mtaup$ values;
denote these contributions as $S'_4,T'_4,U'_4$. Since the
observables ${\cal O}$ are linear functions of $S',T',U'$, we can then
compute ${\cal O}_i(\ng=4)\equiv {\cal O}_i(S'_4,T'_4,U'_4)$.
The predicted values of ${\cal O}_i(\ng=4)$ are then compared to
the best fit experimental values as obtained
using ZFITTER (relative to the reference SM assumed)
and an overall $\Delta \chi^2$ computed.
This computation is performed for each
point of interest in the fourth-generation parameter space.
As noted earlier, we use a reference SM with $\mt(\mt)=165\gev$ and
$m_H=100\gev$ and take $\alpha^{-1}_{em}(m_Z)=129.08$
\refmark{\Swartz}. We employ the latest
values for the following measurements: $m_W/m_Z, \Gamma_Z, \sigma^0_h$,
$R_l, R_b, R_c, A^{0,l}_{FB}$, $A^{0,b}_{FB}, A^{0,c}_{FB}, A_\tau,A_e$
from LEP and $A^0_{LR}$ from SLD.\refmark\data\ Finally, because of
sensitivity of the reference SM prediction from ZFITTER for these
observables to $\alpha_s$, we give results
for the two values, $\alpha_3(m_Z)=(0.12,0.13)$.

\FIG\precisionew{We show the 95\% C.L. and 90\% C.L.
constraints on the $\mbp(\mbp),\mtp(\mtp)$ parameter space (assuming
$\mtaup=\mnup=50\gev$ and $\tanb=1.5$) from $Z$-pole
precision electroweak measurements (coupled with
$\mw/\mz$). We consider two reference SM's specified by $\mt(\mt)=165\gev$,
$m_H=100\gev$,  and the two values,
$\alpha_s(\mz)=0.12$ (dashed lines) and $0.13$
(solid lines), see text.}

The solid and dashed lines in Fig.~\precisionew\
summarize the final results of
this global fit. There, we have adopted $\mtaup=\mnup=50\gev$,
$\tanb=1.5$ and have allowed $\mtp(\mtp)$ and $\mbp(\mbp)$ to vary.
The region in the $\mtp,\mbp$ parameter space {\it above}
the lower (upper) solid line is excluded at $90\% (95\%)$ C.L.
if the reference SM has $\alphaz=0.13$. The two dashed lines give the
same C.L. boundaries in the case $\alphaz=0.12$;
in this latter case, Fig.~\precisionew\ shows that
there is a region at small $\mtp$ that is also excluded.
As one might have naively anticipated,
the region of large $\mtp$ compared to $\mbp$ has a poor C.L.;
however, we also see that if both $\mbp$ and $\mtp$ are small,
then predictions of the fourth-generation model
for $S^\prime,T^\prime,U^\prime$ can again deviate significantly
from the experimentally preferred values, depending upon $\alpha_s(\mz)$.
The region of modest $\mtp(\mtp)\sim\mbp(\mbp)\sim 100\gev$, preferred
on the basis of current and anticipated Tevatron results,
lies well within even the 90\% C.L. limits.

\bigskip
\leftline{\it 7c. Influence on jet and di-jet spectra of a slowly running
four-family $\alpha_s$}

One other indirect effect of the presence of
a fourth family, along with its full complement of supersymmetric
partners, all at relatively low energy scales, has recurred throughout
our discussion: $\alpha_s(Q)$ decreases much more slowly with increasing
$Q$ than in a standard three-family model, especially if the comparison
is with the case where the masses of the superpartners
of the three-family model are taken to be large.  Thus, it is
important to not
forget that $\alpha_s(Q)$ is, in fact, directly measurable through
the single-jet and di-jet spectra of light quark jets. Of course,
the exact scale $Q$ at which $\alpha_s$ is evaluated will be an
important issue.  In the $\msbar$ scheme, higher order calculations
suggest that an appropriate $Q$ value for single-jet inclusive
spectra and di-jet mass spectra is the subprocess center-of-mass energy,
roughly given by $2E_T$ and $m_{2j}$, respectively.
How large can the deviations be?  To illustrate, we consider as a function
of $Q$ the fractional enhancement ratio
$$R_{\alpha_s}\equiv {[\alpha_s(\ng=4)]^2-[\alpha_s^{SM}(\ng=3)]^2 \over
[\alpha_s^{SM}(\ng=3)]^2}\,,\eqn\ralpha$$
where $\alpha_s^{SM}$ is the $\alpha_s$ predicted in the three-family
case with no superpartner effects,
(\ie\ with all superparticle masses taken to be large); results for
$\alpha_s(\ng=4)$ will be illustrated for $\mbp(pole)=\mtp(pole)=105\gev$
(our preferred mass range) and $\msusycol$ values of $200\gev$
and $400\gev$. Such values are quite typical of the
overall mass scales for the strongly-interacting sparticles
(that contribute to $\alpha_s$ evolution) in the model scenarios
we have studied.\foot{Note that $\msusycol$
is not generally the same as the effective $\msusy$ appropriate
in discussing gauge coupling unification, which is sensitive to
the entire particle content of the theory and not just the strongly
interacting particles.}
We employ $\alpha_s^2$ in Eq.~\ralpha\
because this is what appears in the QCD subprocess cross sections.

The results appear in
\FIG\ralphasfig{We plot the percentage
enhancement $R_{\alpha_s}$ (see text) as a function of $Q$
for $\ng=4$ choices of $\msusycol=200\gev$ and $\msusycol=400\gev$.}
Fig.~\ralphasfig.  We see that for $\msusycol=200\gev$, percentage
deviations are, for example, about 25\% at $Q=1\tev$.
This would correspond to subprocess energies of $1\tev$
for the standard $\msbar$ scale choice mentioned above.
The corresponding value
for the (probably more reasonable) choice of $\msusycol=400\gev$ is 12\%.
If there are only three families {\it with}
superpartners at a light mass scale, the values of $R_{\alpha_s}$,
Eq.~\ralpha ( with $\alpha_s(\ng=3)$ in place
of $\alpha_s(\ng=4)$) are very much smaller in all cases.
Thus, it is not entirely impossible that preliminary
observations of this type of deviation by CDF and D0\Ref\deviations{Figures
available on the CDF WWW Home Page and private communications.}
could be a hint that four generations with relatively
light superpartners are present.

\vskip 1in
\bigskip

\leftline{\bf 8. Conclusions}

We have shown that a four-generation MSSM model is an attractive
extension of the usual three-generation MSSM in that gauge unification
and automatic electroweak symmetry breaking via RGE renormalization
both occur naturally for typical soft-SUSY-parameter boundary conditions
at $\mgut$.  {\it A significant, but restricted range of $\tp$ and $\bp$
masses is allowed (see below) even after requiring
that all Yukawa couplings remain
in the perturbative domain throughout $\mz$ to $\mgut$ evolution; however,
Yukawa unification is generally not possible.} The domain of
allowed $\tp$ and $\bp$
masses is impacted by constraints coming from
the SUSY sparticle sector, the most important of which is the
requirement that the lighter $\staup$ eigenstate ($\staupone$)
be more massive than the LEP-I limit of $45\gev$ and
also heavier than the $\cnone$ LSP. This constraint
can rule out a range of lighter $\bp$ masses; the smaller the
soft-SUSY-breaking $m_0$ and $\mhalf$ parameters, the larger the range.
For given $\tp$ and $\bp$
masses the $\staupone$ constraint places significant restrictions
on $m_0$ and $\mhalf$, and thence
on other sparticle masses. Generally, small $m_0$ values
are disfavored unless $\mgl$ is large, and there is a significant
lower bound on $\mgl$ such that no $m_0,A$ choices yield a satisfactory
$\staupone$ mass. There is no analogous bound in the three-generation case.
{\it Thus, the sparticle mass scale must be larger in four- {\it vs.}
three-generation MSSM models.}

Significant differences are also found for the relations
between different sparticle masses in comparing $\ng=4$ results to
those for $\ng=3$. The most dramatic difference arises directly from
the roughly factor of two larger value of the unified gauge coupling
$\alpha(\mgut)$ in the $\ng=4$ case. This leads to ratios such
as $\mslepr/\mcnone$ and $\mslepl/\mcnone$ being approximately a factor
of two larger for $\ng=4$ compared to $\ng=3$ for a given choice
of $m_0/\mhalf$. Indirect tests/verifications
of the presence of a fourth generation
by observing only a light gaugino and sparticles belonging to the first two
families are thus possible.

Direct experimental constraints on the perturbative four-generation MSSM
model are becoming very strong, and near-term experiments could easily
eliminate the model altogether. The $\nup$ and $\taup$ have masses that are
strongly bounded from above for all $\mtp,\mbp>45\gev$ (the current
LEP-I bound) and will
be readily seen at LEP-176. For $\nup$ and $\taup$ masses
just above the current LEP lower bounds (we take $50\gev$ for these masses)
perturbativity constraints force $\mtp(pole)\lsim 160\gev$
and $\mbp(pole)\lsim 120\gev$. We also can be quite certain that
the $\tp$ is heavier than the $\bp$ since otherwise
the $\tp$ decays to $bW$ (or $bW^*$), and $\tp\bar\tp$ final states
would have led
to top-quark-like events at a rate inconsistent with CDF and D0 results.
For $\mtp(pole)>\mbp(pole)$, the $\bp$ can (and must, if
it is to have evaded detection to date) decay via flavor
changing neutral current processes to $b+X$ ($X=\gamma,g,q\anti
q,Z^{(*)},\hl,\ldots$). For expected branching ratios, non-observation of
$\bp\bar\bp$ production events at the Tevatron in which both
$\bp$'s decay to $b+\gamma$ probably excludes $\mbp(pole)\lsim 80-85\gev$.
The range $\mbp(pole)>\mz+\mb(pole)+\sim 5\gev$ (for which $\bp\rta b Z$
decays have significant branching ratio) is probably excluded
by the non-observation of the extra $Z$'s expected from $\bp\bar\bp$ production
events, except for soft-SUSY-breaking parameter scenarios
such that $\mhl$ is small enough for $\bp\rta b\hl$ decays
to dominate over $\bp\rta bZ$ --- the scenarios for which the
$\hl$ channel can be open have a large soft-SUSY-breaking
scalar mass, $m_0$. The top quark searches of CDF
and D0 would have detected production of $\tp\bar\tp$ with $\tp\rta \bp W$
(or $W^*$) and $\bp$ decaying hadronically, unless $\mtp(pole)$
is quite close to $\mbp(pole)$ (implying soft $W^*$ decay products
that evade their hard cuts).
We roughly estimate that $\mtp(pole)$ must be $\lsim 115-120\gev$
on this basis.  All of this leads to a highly preferred mass range for
the $\bp$ and $\tp$:
$$120\gev\gsim \mtp(pole)\gsim \mbp(pole)~~~{\rm with}~~~~
100\gev\gsim\mbp(pole)\gsim 80\gev\,.\eqn\bestregion$$
We estimate that, for an integrated luminosity of $L=150\pbi$,
it is well within the capability
of the Tevatron to either exclude all reasonable allowed models
or detect a signal.

Of course, some of the above restrictions follow from
the assumption that the `top-quark'
events at CDF and D0 arise from the third-generation top.
We have noted that the alternative case (emphasized in Ref.~[\habertop])
where these are
$\tp\bar\tp$ events in which $\tp\rta b W$, while the normal $t$
has $\mt\sim \mw$ and decays via the difficult-to-detect
$t\rta \wt t_1 \cnone$ channel is not possible
in the SUSY GUT context {\it assuming universal $m_0$ and $A$ parameters}.
As one adjusts parameters so as to minimize $m_{\wt t_1}+\mcnone$,
the $\staupone$ always becomes too light, its mass falling
below the LEP-I bound or below the LSP mass, or the theory becomes
inconsistent by virtue of the CP-odd Higgs or some colored and/or charged
4th-family sparticle being required to have $m^2<0$,
long before the lighter $\wt t_1$ reaches masses below $\mw$.

Certainly, if there is a fourth generation with Yukawa couplings
that remain perturbative up to $\mgut$, experimentalists will
discover a plethora of new signals at LEP-II and with increased
luminosity at the Tevatron.
If in the end no signal is found, requiring (roughly) $\mtaup,\mnup\gsim
95\gev$
and $\mtp,\mbp\gsim 200\gev$, determined
four-generation model builders must
become resigned to having one or more of the Yukawa coupling
constants becoming non-perturbative before evolution up
to $\mgut$ is complete.  While there is no known fundamental
reason to disallow this, such a scenario is
distinctly less predictive, and therefore less attractive,
than full perturbative evolution
up to the unification scale for all parameters.

\bigskip

\centerline{\bf Acknowledgments}

This work has been supported in part by
Department of Energy grants \#DE-FG03-91ER40674, \#DE-FG02-85ER40214,
and by the Davis Institute for High Energy Physics.

\appendix

\bigskip

\REF\mv{S. Martin and M. Vaughn, \prdj{50}, 2282 (1994).}
In this Appendix we collect all of the renormalization group equations
for the four-generation extension of the MSSM that are used in our
calculations.  Among the many references available, we primarily
employed Bjorkman and Jones, Ref.~[\Nanojones], Martin and Vaughn,\refmark\mv\
and Cvetic and Preitschopf\refmark\mcvetic\
for the various beta-functions required.

\noindent
Assuming no KM mixing and universal scalar masses at $\mgut$, and defining
$t=$ $({1\over 2\pi})log[Q(\gev)]$, $y_i=\lambda^2_i/(4\pi)$, and
$\alpha_i=g^2_i/(4\pi)$, we summarize the RGE's below. We denote the
anomalous--dimension contributions to ${dy_i\over dt}$ by
$4\pi\gamma_i^{(1)}$ at one loop and $(4\pi)^2\gamma_i^{(2)}$ at two loops
for the $i^{th}$ (unmixed) fermion.

For the Yukawa couplings we write
$${dy_i\over dt}=y_i(4\pi\gamma_i^{(1)}+(4\pi)^2\gamma_i^{(2)})\; ,
\eqn\dydt$$
where $i=t,b,\tau$, $\tpr ,\bpr ,\nupr,\taupr$.
The expressions for $4\pi\gamma_i^{(1)}$
and $(4\pi)^2\gamma_i^{(2)}$ are as follows:

$$\eqalign{
4\pi\gamma_t^{(1)}&=-{13\over15}\alpha_1-3\alpha_2-{16\over
3}\alpha_3 + 6y_t+3y_{\tpr }+y_b+y_{\nupr }\cr
(4\pi)^2\gamma_t^{(2)}&= {3523\over 450}\alpha_1^2 +{27\over
2}\alpha_2^2+{80\over 9}\alpha_3^2+\alpha_1\alpha_2+{136\over
45}\alpha_1\alpha_3\cr
&+8\alpha_2\alpha_3 +16y_t\alpha_3+16y_{\tpr }\alpha_3+{6\over
5}y_t\alpha_1+{4\over 5}y_{\tpr }\alpha_1\cr
&+{2\over 5}y_b\alpha_1+6y_t\alpha_2-22y^2_t-
9y_{\tpr }^2-3y^2_{\nupr }-5y_by_t\cr
&-3y_{\bpr }y_{\tpr }-y_{\taupr }y_{\nupr }-
5y^2_b-9y_ty_{\tpr }-3y_by_{\bpr }-y_by_\tau\cr
&-y_by_{\taupr }-3y_ty_{\nupr }\,;\cr}
\eqn\eqgamt$$

$$\eqalign{
4\pi\gamma_b^{(1)}&=-{7\over 15}\alpha_1-3\alpha_2-{16\over
3}\alpha_3+6y_b+3y_{\bpr }+y_t+y_\tau+y_{\taupr }\cr
(4\pi)^2\gamma_b^{(2)}&={371\over 90}\alpha_1^2 +{27\over 2}\alpha_2^2+{80\over
9} \alpha_3^2+\alpha_1\alpha_2+{8\over 9}\alpha_1\alpha_3\cr
&+8\alpha_2\alpha_3+16y_b\alpha_3+16y_{\bpr }\alpha_3+{2\over 5}
y_b\alpha_1+{4\over 5}y_t\alpha_1\cr
&-{2\over 5}y_{\bpr }\alpha_1+{6\over 5}y_\tau\alpha_1+{6\over
5}y_{\taupr }\alpha_1+6y_b\alpha_2\cr
&-22
y^2_b-9y^2_{\bpr }-3y_\tau^2-3y^2_{\taupr }-5y_by_t-3y_{\bpr }y_{\tpr }-
y_{\taupr }y_{\nupr }\cr
&-5y^2_t-9y_by_{\bpr }-3y_ty_{\tpr }-3y_by_\tau
-3y_by_{\taupr }-y_ty_{\nupr }\,;\cr}
\eqn\eqgamb
$$

$$\eqalign{
4\pi\gamma_\tau^{(1)}&=-{9\over
5}\alpha_1-3\alpha_2+3y_b+3y_{\bpr }+4y_\tau+y_{\taupr }\cr
(4\pi)^2\gamma_\tau^{(2)}&={171\over 10}\alpha_1^2+{27\over
2}\alpha_2^2+{9\over
5}\alpha_1\alpha_2+16y_b\alpha_3+16y_{\bpr }\alpha_3\cr
&-{2\over 5}y_b\alpha_1-{2\over 5}y_{\bpr }\alpha_1 +{6\over 5}y_\tau
\alpha_1+{6\over5}y_{\taupr }\alpha_1+6y_\tau\alpha_2\cr
&-9y_b^2-9y^2_{\bpr }-10y_\tau^2-3y^2_{\taupr }-
3y_by_t-3y_{\bpr }y_{\tpr }-y_{\taupr }y_{\nupr }\cr
&-3y_\tau y_{\taupr }-9 y_\tau y_b-9y_\tau y_{\bpr }\,;\cr}
\eqn\eqgamtau$$

$$\eqalign{4\pi\gamma_{\tpr }^{(1)}&=-{13\over 15}\alpha_1-3\alpha_2-{16\over
3}\alpha_3+6y_{\tpr }+3y_t+y_{\bpr }+y_{\nupr }\cr
(4\pi)^2\gamma_{\tpr }^{(2)}&= {3523\over 450}\alpha_1^2 +{27\over
2}\alpha_2^2+{80\over 9}\alpha_3^2+\alpha_1\alpha_2+{136\over
45}\alpha_1\alpha_3\cr
&+8\alpha_2\alpha_3 +16y_{\tpr }\alpha_3+16y_{t}\alpha_3+{6\over
5}y_{\tpr }\alpha_1+{4\over 5}y_{t}\alpha_1\cr
&+{2\over 5}y_{\bpr } \alpha_1+6y_{\tpr } \alpha_2 -
22 y^2_{\tpr }- 9 y_{t}^2 - 3
y^2_{\nupr }-5y_{\bpr }y_{\tpr }\cr
&-3y_{b}y_{t}-y_{\taupr }y_{\nupr }-5y^2_{\bpr }-
9y_{\tpr }y_{t}-3y_{\bpr }y_b-y_{\bpr }
y_{\taupr }
-y_{\bpr }y_{\tau}-3y_{\tpr }y_{\nupr }\,;\cr}
\eqn\eqgamtp$$

$$\eqalign{
4\pi\gamma_{\bpr }^{(1)}&=-{7\over 15}\alpha_1-3\alpha_2-{16\over
3}\alpha_3+6y_{\bpr }+3y_{b}+y_{\tpr }+y_\tau+y_{\taupr }\cr
(4\pi)^2\gamma_{\bpr }^{(2)}&={371\over 90}\alpha_1^2 +{27\over 2}\alpha_2^2+
{80\over
9} \alpha_3^2+\alpha_1\alpha_2+{8\over 9}\alpha_1\alpha_3\cr
&+8\alpha_2\alpha_3+16y_{\bpr }\alpha_3+16y_{b}\alpha_3+{2\over 5}
y_{\bpr }\alpha_1+{4\over 5}y_{\tpr }\alpha_1\cr
&-{2\over 5}y_{b}\alpha_1+{6\over 5}y_{\taupr }\alpha_1+{6\over
5}y_{\tau}\alpha_1+6y_{\bpr }\alpha_2\cr
&-22y^2_{\bpr }-
9y^2_{b}-3y_{\taupr }^2-3y^2_{\tau}-5y_{\bpr }y_{\tpr }-3y_{b}y_{t}\cr
&-y_{\taupr }y_{\nupr }-5y^2_{\tpr }-9y_{\bpr }y_{b}-
3y_ty_{\tpr }-3y_{\bpr }y_\tau
-3y_{\bpr }y_{\taupr }-y_{\tpr }y_{\nupr }\,;\cr}
\eqn\eqgambp$$

$$\eqalign{
4\pi\gamma_{\taupr }^{(1)}&=-{9\over
5}\alpha_1-3\alpha_2+3y_{\bpr }+3y_{b}+y_\tau+4y_{\taupr }+y_{\nupr }\cr
(4\pi)^2\gamma_{\taupr }^{(2)}&={171\over 10}\alpha_1^2+{27\over 2}\alpha_2^2+
{9\over
5}\alpha_1\alpha_2+16y_{\bpr }\alpha_3+16y_{b}\alpha_3\cr
&-{2\over 5}y_{\bpr }\alpha_1-{2\over 5}y_{b}\alpha_1 +{6\over 5}y_{\taupr }
\alpha_1+{6\over5}y_{\tau}\alpha_1+6y_{\taupr }\alpha_2\cr
&-9y_{\bpr }^2-9y^2_{b}-10y_{\taupr }^2-3y^2_{\tau}-
3y_{\bpr }y_{\tpr }-3y_{b}y_{t}
\cr
&-y_{\taupr }y_{\nupr }-3y_\tau y_{\taupr }-9 y_{\taupr } y_{\bpr }-9y_{\taupr
}
y_{b}-3y_{\nupr }\cr
&-3y_{\nupr }y_t-3_{\nupr }y_{\tpr }-2y_{\taupr }y_{\nupr }\,;\cr}
\eqn\eqgamtaup$$

$$\eqalign{
4\pi\gamma_{\nupr }^{(1)}&=-{3\over
5}\alpha_1-3\alpha_2+3y_{\tpr }+3y_t+4y_{\nupr }+y_{\taupr }\cr
(4\pi)^2\gamma_{\nupr }^{(2)}&={267\over 50}\alpha_1^2+{27\over
2}\alpha_2^2+{9\over 5}\alpha_1\alpha_2 +16y_t\alpha_3+16 y_{\tpr }\alpha_3
+{6\over 5}y_{\taupr }\alpha_1+ 6y_{\nupr }\alpha_2\cr
&+{6\over 5}y_{\nupr }\alpha_1+{4\over 5}y_t\alpha_1 +{4\over 5}y_{\tpr
}\alpha_1-9
y^2_t-9y^2_{\tpr }-10 y^2_{\nupr }-3y_by_t-3 y_{\bpr }y_{\tpr }\cr
&-3y_{\taupr }y_{\nupr }-3y^2_{\taupr }-y_\tau y_{\taupr }-3y_{\taupr }y_b-3
y_{\taupr }y_{\bpr }-9y_{\nupr }y_t\cr
&-9y_{\nupr }y_{\tpr }\,.\cr}
\eqn\eqgamnup
$$
\noindent
For the Higgs mass terms, we have:

$${dm^2_{H_2}\over {dt}}={3 \sum_{q=t,\tpr }y_q
{\cal U}_q+y_{\nupr } {\cal U}_{\nupr }-{3\over 5}
\alpha_1 M^2_1-3 \alpha_2 M^2_2}\,;
\eqn\eqhtwo$$

$${dm^2_{H_1}\over {dt}}={3 \sum_{q=b,\bpr }y_q
{\cal D}_q+\sum_{l=\tau,\taupr }y_{l} {\cal D}_{l}-{3\over 5}
\alpha_1 M^2_1-3 \alpha_2 M^2_2}\,.
\eqn\eqhone$$

\noindent
The third and fourth generation scalar quark and lepton soft
mass terms evolve according to:
$${dm^2_{Q,Q'}\over {dt}}={ y_{t,\tpr }
{\cal U}_{t,\tpr }+y_{b,\bpr } {\cal D}_{b,\bpr }-{1\over 15} \alpha_1 M^2_1-
3 \alpha_2 M^2_2-{16 \over 3} \alpha_3 M^2_3}\,;
\eqn\eqmq$$
$${dm^2_{U,U'}\over {dt}}={2 y_{t,\tpr }
{\cal U}_{t,\tpr }-{16\over 15} \alpha_1 M^2_1-
{16 \over 3} \alpha_3 M^2_3}\,;
\eqn\eqmu$$
$${dm^2_{D,D'}\over {dt}}={2 y_{b,\bpr } {\cal D}_{b,\bpr }-{4\over 15}
\alpha_1 M^2_1-{16 \over 3} \alpha_3 M^2_3}\,;
\eqn\eqmd$$
$${dm^2_{L,L'}\over {dt}}={y_{\tau,\taupr }
{\cal D}_{\tau,\taupr }+\delta_{i L'}y_{\nupr } {\cal U}_{\nupr }-
{3\over 5} \alpha_1 M^2_1-3 \alpha_2 M^2_2}\,;
\eqn\eqml$$
$${dm^2_{E,E'}\over {dt}}={2 y_{\tau,\taupr }
{\cal D}_{\tau,\taupr }-{12\over 5} \alpha_1 M^2_1};\;{dm^2_{N}\over {dt}}=
{2 y_{\nupr }{\cal U}_{\nupr }}\,.
\eqn\eqme$$

\noindent
In the above,
$${\cal U}_{t,\tpr }=m^2_{H_2}+m^2_{Q,Q'}+m^2_{U,U'}+A^2_{t,\tpr },
\eqn\equt$$
$${\cal U}_{\nupr }=m^2_{H_2}+m^2_{L'}+m^2_{N}+A^2_{\nupr },
\eqn\equnu$$
$${\cal D}_{b,\bpr }=m^2_{H_1}+m^2_{Q,Q'}+m^2_{D,D'}+A^2_{b,\bpr },
\eqn\eqdb$$
$${\cal D}_{\tau,\taupr }=m^2_{H_1}+m^2_{L,L'}+m^2_{E,E'}+A^2_{\tau,\taupr }.
\eqn\eqdtau$$

\noindent
The running of the first and second generation scalar quark and lepton
soft-SUSY-breaking masses is obtained
by neglecting all Yukawa couplings in the expressions above.
The parameters $\mu$, specifying mixing of the Higgs superfields
in the superpotential, and $B$, such that $B\mu$ is the coefficient
of the $H_1H_2$ scalar field mixing term in the soft-SUSY-breaking
potential, evolve according to:
$${dlog(\mu)\over {dt}}={{3\over 2} \sum_{q=t,\tpr ,b,\bpr } y_{q}+
{1\over 2}\sum_{l=\tau,\taupr ,\nupr } y_l-{3\over 10} \alpha_1-
{3\over 2} \alpha_2}\,;
\eqn\eqmu$$
$${dB\over {dt}}={3 \sum_{q=t,\tpr ,b,\bpr } y_{q} A_q+
\sum_{l=\tau,\taupr ,\nupr } y_{l} A_l+{3\over 5} \alpha_1 M_1+
{3} \alpha_2 M_2}\,.
\eqn\eqbterm$$

\noindent
The running of the third and fourth family
soft-SUSY-breaking potential tri-linear
term $A$ coefficients is given by:
$$
\eqalign{{dA_{t,\tpr }\over {dt}}=& 6 A_{t,\tpr }y_{t,\tpr }+
A_{b,\bpr }y_{b,\bpr }+3 A_{\tpr ,t}y_{\tpr ,t}+A_{\nupr }y_{\nupr }\cr
&+{13\over 15} \alpha_1 M_1+3 \alpha_2 M_2+{16\over 3} \alpha_3 M_3\,;\cr}
\eqn\eqat$$
$$\eqalign{{dA_{b,\bpr }\over {dt}}=& 6 A_{b,\bpr }y_{b,\bpr }+
A_{t,\tpr }y_{t,\tpr }+3 A_{\bpr ,b}y_{\bpr ,b}+
\sum_{l=\tau,\taupr }A_{l}y_{l}\cr
&+{7\over 15} \alpha_1 M_1+3 \alpha_2 M_2+{16\over 3} \alpha_3 M_3\,;\cr}
\eqn\eqab
$$
$$
\eqalign{{dA_{\tau,\taupr }\over {dt}}=& 4 A_{\tau,\taupr }y_{\tau,\taupr }+
A_{\taupr ,\tau}y_{\taupr ,\tau}+3 \sum_{q=b,\bpr }A_{q}y_{q}+
\delta_{i,{\taupr }}A_{\nupr }y_{\nupr }\cr
&+{9\over 5} \alpha_1 M_1+{3} \alpha_2 M_2\,;\cr}
\eqn\eqatau
$$
$$
{dA_{\nupr }\over {dt}}={4 A_{\nupr }y_{\nupr }+
A_{\taupr }y_{\taupr }+3 \sum_{q=t,\tpr }A_{q}y_{q}
+{3} \alpha_2 M_2+{3\over 5} \alpha_1 M_1}\,.
\eqn\eqanu
$$

The Higgs boson, scalar quark and scalar lepton mass--squared RGE's generally
include a term in each equation that is a numerical factor times $\alpha_1{\cal
S}',$ where
$$
\eqalign{{\cal S}'&\equiv m^2_{H_2}-m^2_{H_1}+m_Q^2+m^2_{Q'}-2m^2_U+m^2_D\cr
&-2m^2_{U'}+m^2_{D'}-m_L^2-m_{L'}^2+m^2_E+m^2_{E'}\; .\cr}
\eqn\eqsprimedef
$$
Substituting from the RGE's with ${\cal S}'$ terms included, one finds
$${d{\cal S}'\over dt}={46\over 5}\alpha_1{\cal S}'\; .
\eqn\eqsprime
$$
Because of the universal scalar mass boundary
condition at $\mgut$ that we assume, ${\cal S}'(\mgut)=0$. Therefore ${\cal
S}'(t)=0$ for all $t$ and we do not include this term in the
mass--squared RGE's presented above.

The final soft-SUSY-breaking parameters are the gaugino masses.
Using the notation $\ng$ for the number of generations,
these evolve according to:
$$
\eqalign{
{d M_1\over dt}=&\bigl(2\ng+{3\over 5}\bigr) \alpha_1 M_1\,;\cr
{d M_2\over dt}=&\bigl(2\ng-5\bigr) \alpha_2 M_2\,;\cr
{d M_3\over dt}=&\bigl(2\ng-9\bigr) \alpha_3 M_3\,.\cr}
\eqn\eqgaugino
$$

Finally, we present the gauge coupling RGE's for completeness. These are the
same as the massless fourth generation neutrino case, since the extra right
handed neutrino that provides a mass term in our case is decoupled from the
gauge sector. Keeping the number of generations, $\ng$, explicit,
including the third and fourth (unmixed) generation Yukawa couplings,
and including exactly two Higgs doublets (as appropriate
for the MSSM), one has the gauge RGE's:
$$
\eqalign{{d\alpha_1\over dt}&=\biggl(2\ng+{3\over 5}\biggr)\alpha_1^2
+{\alpha_1^2\over
4\pi}\bigg\lbrack \biggl({38\over 15}\ng+{9\over
25}\biggr)\alpha_1+\biggl({6\over 5}\ng+{9\over 5}\biggr)\alpha_2+{88\over
15}\ng\alpha_3\cr
&-{18\over 5}(y_\tau+y_{\taupr })-{14\over 5}(y_b+y_{\bpr })-{26\over
5}(y_t+y_{\tpr })\bigg\rbrack\,;\cr}
\eqn\eqalphaone
$$
$$
\eqalign{{d\alpha_2\over dt}=(2\ng-5)\alpha_2^2&+{\alpha_2^2\over
4\pi}\bigg\lbrack \biggl({2 \over 5}\ng +{3\over 5}\biggr)\alpha_1 +
(14\ng-17)\alpha_2+8\ng\alpha_3\cr
&-2(y_\tau+y_{\taupr })-6(y_b+y_{\bpr }+y_t+y_{\tpr })\bigg\rbrack\,;\cr}
\eqn\eqalphatwo
$$
$$
\eqalign{{d\alpha_3\over dt}=(2\ng-9)\alpha_3^2&+{\alpha_3^2\over
4\pi}\bigg\lbrack {11\over 15}\ng\alpha_1+3\ng\alpha_2 +
\biggl({68\over 3}\ng-54\biggr)\alpha_3\cr
&-4(y_b+y_{\bpr}+y_t+y_{\tpr })\bigg\rbrack\; .\cr}
\eqn\eqalphathree
$$

The renormalization group equations were implemented numerically, and
iteration was employed to find a fully consistent solution for the
complete evolution between $\mgut$ and $\mz$.

\bigskip
\appendix

In this appendix we discuss the question of the accuracy with
which $\cos2\beta$ and, thence, $\tanb$, can be determined
using the mass differences of Eq.~\deli.\Ref\murayamapc{We
thank H. Murayama for several conversations on various aspects
of the discussion in this Appendix. See also Ref.~\dpfsusy.}

To use $\mslepl^2-\msnu^2$ requires accurate
determinations of $\msnu$ and $\mslepl$.
The required measurements are best performed at an $\epem$ collider
where the center-of-mass energy is precisely known and the energy
spectra of final leptons can then be directly related to
the masses of the produced particles and their decay products.
Determination of $\mslepl$ is relatively straightforward
since $\slepl\rta l\cnone$ is almost always an important, if not dominant,
mode --- if $\mslepl>\mcpone\sim\mcntwo$, then $\slepl\rta \cpone \nu$
and $\slepl\rta \cntwo\ell$ can be competitive. A fit to the $\ell$
spectrum will yield the masses of both the $\slepl$ and the $\cnone$.
Should the $\cpone\nu$ and $\cntwo\ell$ modes also be important,
the fit would presumably reveal the presence of several upper end points,
and allow determination of both $\mcnone$ and $\mcntwo$.  A Monte Carlo
study is desirable to determine the exact accuracy with which
all masses could be determined in the multiple decay case.
Perhaps accuracies of about $\pm 10\gev$ could be achieved.
For $\mslepl\sim 240\gev$ (near our lower limit in the $\ng=4$ case
scenarios, but already rather marginal for a $\sqrt s=500\gev$ NLC)
this would correspond to accuracies of $\sim4\%$, \ie\
on the edge of what would be useful.

Determination of $\msnu$ follows a similar pattern. We give below
discussions for two scenarios:
a) $\msnu<\mcpone$ and b) $\msnu>\mcpone$. Scenario a) does
not arise for the specific $\ng=4$ models illustrated in
Figs.~\spectrumdilaton-\spectrumhighm, but might be relevant if there
was significant non-universality in the $\mgut$-scale soft-SUSY-breaking
parameters.  For $\ng=4$, universal $m_0$ and $M_i$ values
always lead to $\msnu>\mcpone$, as shown.

If $\msnu<\mcpone$ then the $\cpone$ decays to a mixture
of $\slepl \nu$ and $\ell\snu$; since $\mslepl>\msnu$ the latter
decay is usually the dominant one --- indeed, in many
cases $\mslepl-\msnu$ is large enough that the $\cpone\rta\slepl\nu$ channel
is forbidden even though the $\cpone\rta\ell\snu$ channel is open.
At an $\epem$ collider
the end-points of the $\ell$ spectrum from $\cpone\cmone$
production followed by $\cpmone\rta\ell\snu$ decay provide a determination
of both $\mcpone$ and $\msnu$;
direct observation of the $\snu$ is not required, a fortunate
fact given that it decays invisibly to $\nu\cnone$.

If $\msnu>\mcpone$, then $\snu\rta \cpone \ell$, and the $\cpone$
decays via a virtual sneutrino or slepton to
the three-body mode $\ell\nu\cnone$, via virtual squark to $q\anti q\cnone$,
or via virtual or real $W$ to a mixture of these two final states.
Analogous to the previous case, the $\ell$ spectrum end-points
will allow determination of $\msnu$ and $\mcpone$, although one will
have to carefully account for the underlying
smooth $\ell$ spectrum from the $\cpone$
decay if events containing only $2\ell+4j+$missing energy (which avoid
contamination from the soft $\ell$(s) from $\cpmone$ decay) do not
occur at a sufficient rate. Ideally, it would be good to learn first
about the $\cpone\cmone$ channel by setting $\sqrt s$ to a value
above $2\mcpmone$ but below (or not far above) $2\msnu$ ---
due to the rapid turn-on
of a two-fermion channel, the rate for $\cpone\cmone$ production
might be reasonable,
while the $\snu\snu$ rate could be zero or quite small (due to
a large $\msnu-\mcpone$ splitting and/or
the slow turn-on of the $\snu\snu$ spin-0 pair channel).  To study $\snu\snu$,
one would then up the energy to a level such that the $\snu\snu$
rate was large.

A significant complication is the fact that $\slepl\slepl$
production would also be present at a similar rate
to $\snu\snu$ (recall that
$\mslepl\gsim\msnu$). For the present case of $\msnu>\mcpmone$,
$\slepl\rta \ell\cnone,\nu\cpone,\ell\cntwo$ will all occur,
most probably with similar branching ratios.
Events in which both $\slepl$'s decay to $\ell\cnone$ can be eliminated
by requiring some jet activity (which is generally
present for the $\snu\snu$ final states). If both $\slepl$'s decay to
$\cpmone\nu$, then the only charged $\ell$'s
would be from decays of the $\cpmone$'s,
which would yield a smooth (and relatively soft) $\ell$ spectrum;
in addition, these $\slepl\slepl$ events would tend to have larger missing
energy than the $\snu\snu$ events.  If one $\slepl$ decays to $\ell\cnone$
and the other to $\nu\cpone$, followed by $\cpone\rta \ell\nu\cnone$,
then we could end up with a $2\ell+2j+$missing energy final state
that might not be easily distinguished from the $2\ell+4j+$missing
energy state of interest for $\snu\snu$ events (certainly not
all jets in the latter case would be detectable). But, the spectrum
of the $\ell$ from the $\cpone$ decay would be soft and smooth,
so that the $\ell$-spectrum thresholds present in $\snu\rta \ell\cpmone$
might still stand out.

Thus, it would seem that there
are means for isolating the $\snu\snu$ events of interest and
extracting the $\ell$ spectrum thresholds that would allow a reasonable
determination of $\msnu$ (especially if $\mcpmone$ is known from
a lower energy measurement). Unfortunately, these complicated
scenarios have not been explicitly studied for an $\epem$ collider.
The results for simpler scenarios (see Ref.~[\dpfsusy] and
references therein) suggest the crude estimate
that mass determinations for $\msnu$
might be possible within $\pm5\%$, at least
for masses of order $200\gev$ and below ({\it i.e.}
comparable to the minimum possible $\msnu$ values
in the scenarios discussed in the previous sub-section.)
However, including the fact that the $\mslepl$ and $\msnu$ errors must
be combined in quadrature, despite this fairly
small uncertainty in the mass measurements we would not have
the accuracy required for a $3\sigma$
sensitivity to $\cos2\beta$, even for the smallest
possible masses allowed in the $\ng=4$ case.

Let us now turn to the $\msdnl^2-\msupl^2$ measurement.
First, there is a very real
possibility (a near certainty for $\ng=4$) that the squarks are simply
so heavy, $\msq\gsim 300\gev$, that impossible accuracy would
be required for sensitivity to $\cos2\beta$.
Even for masses below this level, it is still far from clear
that the required accuracy can be achieved.
Squarks will decay to a quark plus real or virtual
gluino, the latter then decaying to a variety of final states
(with $q\anti q \cnone$ unlikely to be dominant for the
larger $\mgl$ predicted for $\ng=4$).
Jet spectra end-points (which would appear on top of
a smooth jet-spectrum background from the real or virtual
$\gl$ decays) could provide a certain level of
accuracy in the squark mass determination, but there would be a lot
of overlap of the spectra from different $\gl$ decay channels,
and of the spectra from the $\sdn$ and $\sup$ decays themselves, that might
very well prevent a good determination of $\msdn-\msup$.
An extremely careful Monte Carlo study is required to be able to
determine with any certainty the level of accuracy that can be achieved.

Overall, we conclude that
our ability to determine $\cos2\beta$ and thence $\tanb$
must remain a topic of further study.  Fortunately, the uncertainty
in $\cos2\beta$ is not the limiting factor determining
the level of accuracy with which other mass sum rules and relations
can be tested.  Other experimental and theoretical uncertainties
are almost certainly more important.

\refout
\figout
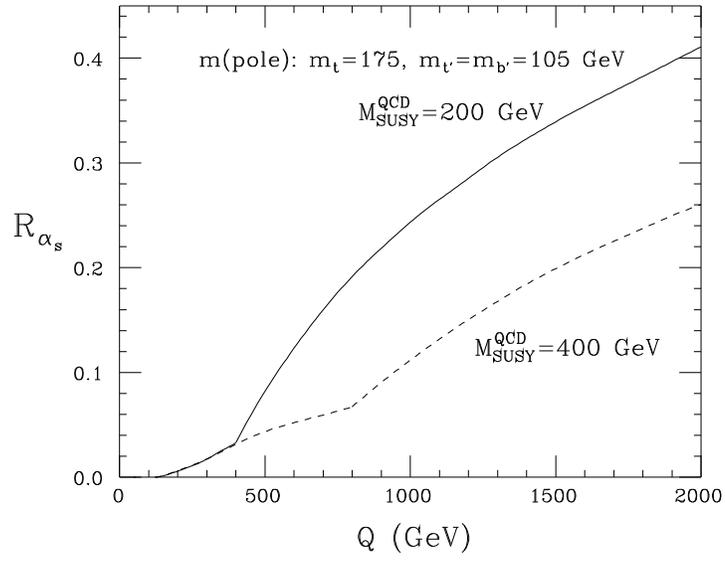
\bye